\documentclass[prd,superscriptaddress,twocolumn,nofootinbib,showpacs,preprintnumbers,floatfix]{revtex4}

\usepackage{mathrsfs}
\usepackage{amsfonts,amssymb,amsmath}
\usepackage{graphicx,epsfig}
\usepackage{multirow}

\begin{document}

\title{The Ultra-High Jet Multiplicity Signal of\\Stringy No-Scale $\cal{F}$-$SU(5)$ at the $\sqrt{s}=7$~TeV LHC}

\author{Tianjun Li}

\affiliation{George P. and Cynthia W. Mitchell Institute for
Fundamental Physics, Texas A$\&$M University, College Station, TX 77843, USA }

\affiliation{Key Laboratory of Frontiers in Theoretical Physics,
Institute of Theoretical Physics, Chinese Academy of Sciences,
Beijing 100190, P. R. China }

\author{James A. Maxin}

\affiliation{George P. and Cynthia W. Mitchell Institute for
Fundamental Physics, Texas A$\&$M University, College Station, TX 77843, USA }

\author{Dimitri V. Nanopoulos}

\affiliation{George P. and Cynthia W. Mitchell Institute for
Fundamental Physics, Texas A$\&$M University, College Station, TX 77843, USA }

\affiliation{Astroparticle Physics Group, Houston Advanced Research Center (HARC),
Mitchell Campus, Woodlands, TX 77381, USA}

\affiliation{Academy of Athens, Division of Natural Sciences,
 28 Panepistimiou Avenue, Athens 10679, Greece }

\author{Joel W. Walker}

\affiliation{Department of Physics, Sam Houston State University,
Huntsville, TX 77341, USA }

%%%%%%%%%%%%%%%%%%%%%%%%%%%%%%%%%%%%%%%%%%%%%%%%%%%%%%%%%%%%%%%%%%%%%%%%%%%%

\begin{abstract}

We present the distinctive collider signatures of No-Scale ${\cal F}$-$SU(5)$, a highly efficient and phenomenologically
favored model built on the tripodal foundations of the $\cal{F}$-lipped $SU(5)\times U(1)_{\rm X}$ Grand Unified Theory,
extra $\cal{F}$-theory derived TeV scale vector-like particle multiplets, and the dynamic high scale boundary
conditions of No-Scale Supergravity.  The identifying features of the supersymmetric spectrum are a light stop and gluino,
with both sparticles much lighter than all the additional squarks.  This unique mass hierarchy leads to the enhanced
production of events with an ultra-high multiplicity of hadronic jets which should be clearly visible to the $\sqrt{s} = 7$~TeV
LHC at only $1~{\rm fb}^{-1}$ of integrated luminosity.  We suggest a modest alternative event cutting procedure based
around a reduced minimal transverse momentum per jet ($p_{\rm T} > 20$~GeV), and an increased minimal multiplicity
($\ge 9$) of distinct jets per subscribed event.
These criteria optimize the $\cal{F}$-$SU(5)$ signal to background ratio,
while readily suppressing the contribution of all Standard Model processes, allowing moreover a clear differentiation
from competing models of new physics, most notably minimal supergravity.
The characteristic No-Scale signature is quite stable across the viable parameter space, modulo an overall rescaling
of the mass spectrum;  Detection by the LHC of the ultra-high jet signal would constitute a suggestive evocation
of the intimately linked stringy origins of $\cal{F}$-$SU(5)$, and could possibly provide a glimpse into the underlying
structure of the fundamental string moduli.

\end{abstract}

\pacs{11.10.Kk, 11.25.Mj, 11.25.-w, 12.60.Jv}

\preprint{ACT-04-11, MIFPA-11-09}

\maketitle

%%%%%%%%%%%%%%%%%%%%%%%%%%%%%%%%%%%%%%%%%%%%%%%%%%%%%%%%%%%%%%%%%%%%%%%%%%%%

\section{Introduction\label{sct:intro}}

The Large Hadron Collider (LHC) at CERN has been steadily accumulating data from 
${\sqrt s}=7$ TeV proton-proton collisions since March 2010. It is expected to reach
an integrated luminosity of $1~{\rm fb}^{-1}$ by the end of 2011, and 
probably $3~{\rm fb}^{-1}$ by the end of 2012, all in search of new physics beyond the
Standard Model (SM).  Supersymmetry (SUSY), which provides a natural
solution to the quantum stability of the gauge hierarchy, is the most promising such
SM extension.  Data corresponding to the paltry integrated luminosity of $35~{\rm pb}^{-1}$
has already been able to establish new constraints on the viable parameter
space~\cite{Khachatryan:2011tk, daCosta:2011hh, daCosta:2011qk}, due to the unprecedented center of mass
collision energy now available.  The search strategy for SUSY signals in the early LHC data has
been actively and eagerly studied by quite a few
groups~\cite{Baer:2010tk,Kane:2011zd, Feldman:2011me, Buchmueller:2011aa, Guchait:2011fb},
with particular focus on the parameter space featuring a traditional mass relationship between squarks
and the gluino, such as a gluino heavier than all squarks or a gluino lighter than all squarks.

A question of great interest is whether 
there exist SUSY models which are well motivated by fundamental theoretical
considerations, for example string derived model building techniques, which can be tested in the initial LHC run.
We consider such a model in this paper, dubbed No-Scale $\cal{F}$-$SU(5)$ ({\it cf.}~Appendix), which traces its lineage directly
from origins as a consistently realized vacuum of F-theory. It is obedient to the strictest bottom-up
phenomenological constraints and maintains non-trivial consistency with equally strict top-down theoretical dynamics.
We arrive ultimately in this work at a detailed elaboration of its distinctive collider level signatures, including
a proposal for modest alterations to the canonical background selection cut strategy which are expected to yield
significantly enhanced resolution of the characteristic ultra-high jet multiplicity $\cal{F}$-$SU(5)$ events.

SUSY represents an intermixing of internal and Poincar\'{e} symmetries, gracefully evading the Coleman-Mandula
theorem via graded extension of the Lie algebra to include anti-commutation.  When localized,
as all fundamental symmetries in string theory must be, the spacetime derivative is therefore
made generally covariant, and SUSY becomes supergravity (SUGRA).  However, not any supergravity
is sufficient. We require cosmological flatness, a highly non-trivial feature which arises
automatically within the context of the 
No-Scale SUGRAs~\cite{Cremmer:1983bf,Ellis:1983sf, Ellis:1983ei, Ellis:1984bm, Lahanas:1986uc}.  
No-Scale SUGRA provides, moreover, an indispensable
mechanism for the dynamic determination of the ``Moduli'' $T_i$, {\it i.e.}, the size and shape of the six-dimensional
compactified space of string theory, thus stabilizing the geometry of our Universe.
\begin{equation}
\frac{d V_{\rm Universe}}{d T_i} = 0 \quad;\quad i = 1,2,\ldots
\label{eq:sns}
\end{equation}
As a key example, the gravitino mass $M_{3/2}$, or by proportional equivalence, the universal gaugino mass
$M_{1/2}$,  is dynamically determined through its explicit dependence on such a modulus.
$M_{1/2}$ is the supersymmetry breaking scale in the simplest No-Scale SUGRAs, which thus 
determines in turn the masses of the supersymmetric particles that are sought at the LHC; The LHC may well
``measure'' the overall size of the compact dimensions, as stabilized by the ``Super No-Scale'' Mechanism
of Eq.~(\ref{eq:sns}), by detecting and studying the SUSY spectrum in some detail.
We have argued~\cite{Li:2011dw} that the extraordinarily large $\mathcal{O}\,(10^{500})$ ``landscape''
of presumptive consistent stringy vacua may be linked by common adherence to the No-Scale principle, as necessary
for some suitably defined notion of energy conservation to apply in the emergence, {\it ex niholo}, of a
cosmologically flat universe with unique locally established moduli from the quantum ``nothingness''.

The vector-like fields which feature essentially in our model trace also a stringy origin, having been  
consistently described within the F-theory model building context~\cite{Jiang:2009zza, Jiang:2009za}.  Their
effect on the renormalization group running of the couplings embedded within the Flipped $SU(5)$
GUT~\cite{Barr:1981qv,Derendinger:1983aj,Antoniadis:1987dx} have been carefully studied, including the
vanishing of the tree level $\beta$-function coefficient of $SU(3)_{\rm C}$.  Most dramatically, this causes
the dual $SU(3)_{\rm C} \times SU(2)_{\rm L}$ and $SU(5)\times U(1)_{\rm X}$ unification scales to become widely
separated, allowing for the gravitational decoupling scenario to be consistently realized.
The distinctively structured vector-like multiplets, which we have named {\it flippons}, with mass
$M_V \simeq 1$~TeV are themselves ultimately testable at LHC, although possibly not during the
initial $7$~TeV run.  There is another prominent consequence of the flatness of the $\alpha_3$ coupling however,
namely the likewise flat tracing of the colored gaugino mass, leading to a conspicuously light gluino, and the
distinctively predictive $m_{\widetilde{t}} < m_{\widetilde{g}} < m_{\widetilde{q}}$ mass hierarchy between the stop, gluino,
and the heavier quark superpartners.  As we shall elaborate, this spectrum, stably characteristic of No-Scale
$\cal{F}$-$SU(5)$, generates a unique event topology due to the $\widetilde{g} \rightarrow \widetilde{t}$ transition,
which will ultimately result in a spectacular signal of ultra-high multiplicity final state jet events.

We will demonstrate that the No-Scale $\cal{F}$-$SU(5)$ scenario can be clearly distinguished from the SM background and
also from the competition of various minimal supergravity (mSUGRA) based benchmarks, and if correct, that it should be visible
to the LHC by the end of 2011~\cite{Li:2011hr}.  Because the supersymmetric particle spectra are quite similar throughout the
previously advertised ``golden strip''~\cite{Li:2010mi} region, and the entire viable parameter space is moreover quite small,
it seems that the model as a whole may be probed by the LHC by the end of 2012.  We emphasize that verification of the distinctive
No-Scale $\cal{F}$-$SU(5)$ signature would also provide a strong indication of the model's string-theoretic heritage.
The No-Scale framework may itself be traced all the way back to the string level property of scale invariance on the
world sheet, a subgroup of the fundamental conformal invariance of the world sheet string action, insomuch as the
vanishing of the two-dimensional $\beta$-function leads to the relation $d V_{\rm eff} / d \phi$ for the
effective scalar potential not only at a single point of the two-dimensional world sheet, but also along an
extended flat direction in four dimensions.  The probing of the SUSY spectrum at LHC may indeed then be
considered a probe of the stringy origin of our Universe, testing a ``string'' of nested assumptions and dependencies,
and possibly even opening a darkened glass upon the hidden workings of the No-Scale Multiverse.

%%%%%%%%%%%%%%%%%%%%%%%%%%%%%%%%%%%%%%%%%%%%%%%%%%%%%%%%%%%%%%%%%%%%%%%%%%%%

\section{Concept To Computer To Collider\label{sct:ccc}}

The gulf separating the theoretical inception from the experimental inquest of a physical model can be quite wide.
Known processes, sufficiently well understood to be relegated to subservience as calibration, will, by their
definition as the easier target, comprise a background which tends to swamp any purported signal of new physics.
The severe synchrotron radiation limits on light particles has forced circular ring collider probes at the energy
frontier to abandon the clean kinematic consumption of elemental electron-positron pairs for the muddled partial
interactions of strongly bound quark-gluon composites.  Any given set of final states, even assuming perfect
efficiency in measurement, and admitting the inevitable evanescence of the neutrino, will correspond to an
innumerably large amalgam of unobservable internal processes.  The statistical variation inherent in quantum
interactions will create false excesses and shortfalls in production which both mask and masquerade as the
sought post Standard Model contributions.

Likewise wide may be the gulf of culture separating the experimental and theoretical communities themselves.
As members of the latter, we have resolutely attempted in our study to appropriate the standard language and tools
of industry of the former in order to facilitate a clear and testable description of the signal which our preferred model
might present at the LHC.  This translation may be logically subdivided into five steps, the first four effected here
via widely established public computer code, and the last accomplished by a program of our own authorship.

For the initial phase of generation of the low order Feynman diagrams which may link the incoming beam to
the desired range of hard scattering intermediate states, we have used the program {\tt MadGraph}~\cite{Stelzer:1994ta}.
All 2-body SUSY processes are included in our simulation.
These diagrams have subsequently been fed into the sister program {\tt MadEvent}~\cite{Alwall:2007st} for appropriate kinematic scaling,
to yield batches of Monte Carlo simulated parton level scattering events.  We implement MLM matching to preclude double counting of final states,
and we use the CTEQ6L1 parton distribution functions to generate the leading Standard Model background.
The cascaded fragmentation and hadronization of these events into final state showers of photons, leptons, and mixed jets
has been handled by {\tt PYTHIA}~\cite{Sjostrand:2006za}.  Finally, a veil of obfuscation must be drawn across the detailed
omniscience of the initial three steps, simulating the limits on information and
vulnerability to error of a physical detector environment; For this fourth processing phase, we have used the program
{\tt PGS4}~\cite{PGS4}, with Level~0 (passive) triggering. We have opted to employ the supplied CMS detector statistics card,
although essentially similar results are expected for the corresponding ATLAS detector card.  These four essential pieces of code have in fact been conveniently
bundled into a highly cohesive operating suite~\cite{MGME}, significantly streamlining the user experience of this sequence of computations, which exhibit,
in actuality, a daunting underlying complexity.

The output produced by the described suite of utilities is not, however, directly suitable for human consumption
or for the efficient discernment of signal from background.  For these purposes, a fifth processing phase is required
to implement the desired cuts, optimized to reduce the background while emphasizing the signal, and to count and compile
the associated net statistics.  We have opted for a proprietary solution in this last step, which we are releasing 
along with the publication of the present document for comparison and reuse by similarly interested research groups.
The Perl script, named {\tt CutLHCO}~\cite{cutlhco} for its operation on the standardized ``.lhco'' {\tt PGS4} output files,
is available for download at the web address given in the citation, and will be described in some detail subsequently,
in Section~\ref{sct:cuts}. 

%%%%%%%%%%%%%%%%%%%%%%%%%%%%%%%%%%%%%%%%%%%%%%%%%%%%%%%%%%%%%%%%%%%%%%%%%%%%

\section{The $\cal{F}$-$SU(5)$ Mass Hierarchy\label{sct:spectrum}}

Application of the dynamically established boundary conditions of No-Scale
Supergravity~\cite{Cremmer:1983bf,Ellis:1983sf, Ellis:1983ei, Ellis:1984bm, Lahanas:1986uc} at the
elevated secondary unification scale of Flipped $SU(5)$ with F-theory derived TeV scale vector-like multiplets establishes
a model which we have named No-Scale $\cal{F}$-$SU(5)$ ({\it cf.}~Appendix).  There is a
highly constrained ``golden''~\cite{Li:2010ws,Li:2010mi} region of parameter space which simultaneously satisfies
all known experimental constraints, moreover featuring an imminently observable proton decay rate~\cite{Li:2010dp,Li:2010rz}.
Our collider simulation uses the spectrum for the ${\cal F}$-$SU(5)$ point in Table~\ref{tab:masses}, featuring a universal gaugino boundary of
$M_{1/2} = 410$ GeV, and a ratio of up- to down-type Higgs vacuum expectation values (VEVs) $\tan \beta = 19.5$.
The LSP neutralino is 99.8\% Bino. Similarly to the mSUGRA picture, our benchmark point is in the stau-neutralino
coannihilation region, but the gluino is lighter than all the squarks except for the light stop in our models.

\begin{table}[htbp]
  \small
	\centering
	\caption{Spectrum (in GeV) for the benchmark point. 
	Here, $M_{1/2}$ = 410 GeV, $\tan \beta = 19.5$, $M_{V}$ = 1 TeV, $m_{t}$ = 174.2 GeV, $M_{Z}$ = 91.187 GeV,
	$\Omega_{\chi}$ = 0.11, $\sigma_{SI} = 3 \times 10^{-10}$ pb. The central prediction for
	the $p \!\rightarrow\! {(e\vert\mu)}^{\!+}\! \pi^0$ proton lifetime is around $5 \times 10^{34}$ years.
	The lightest neutralino is 99.8\% Bino.}
		\begin{tabular}{|c|c||c|c||c|c||c|c||c|c||c|c|} \hline		
    $\widetilde{\chi}_{1}^{0}$&$76$&$\widetilde{\chi}_{1}^{\pm}$&$165$&$\widetilde{e}_{R}$&$157$&$\widetilde{t}_{1}$&$423$&$\widetilde{u}_{R}$&$865$&$m_{h}$&$120.4$\\ \hline
    $\widetilde{\chi}_{2}^{0}$&$165$&$\widetilde{\chi}_{2}^{\pm}$&$756$&$\widetilde{e}_{L}$&$469$&$\widetilde{t}_{2}$&$821$&$\widetilde{u}_{L}$&$939$&$m_{A,H}$&$814$\\ \hline
    
    $\widetilde{\chi}_{3}^{0}$&$752$&$\widetilde{\nu}_{e/\mu}$&$462$&$\widetilde{\tau}_{1}$&$85$&$\widetilde{b}_{1}$&$761$&$\widetilde{d}_{R}$&$900$&$m_{H^{\pm}}$&$820$\\ \hline
    $\widetilde{\chi}_{4}^{0}$&$755$&$\widetilde{\nu}_{\tau}$&$452$&$\widetilde{\tau}_{2}$&$462$&$\widetilde{b}_{2}$&$864$&$\widetilde{d}_{L}$&$942$&$\widetilde{g}$&$561$\\ \hline
		\end{tabular}
		\label{tab:masses}
\end{table}

Due to the stringent No-Scale boundary condition $B_{\mu}=0$ on the soft SUSY breaking coupling from 
the bilinear Higgs mass term $\mu H_d H_u$, the updated ``golden strip''~\cite{Li:2010mi,LMNW-P} has only a small
viable parameter space. In the simplest No-Scale Supergravity models, all the SUSY breaking
soft terms arise from a single parameter $M_{1/2}$, and thus the resulting supersymmetric
particle (sparticle) spectra are structurally similar, modulo a small overall rescaling
of $M_{1/2}$.  The sparticle branching ratios are almost identical across the board. 
For our purposes then, the tabulated point is completely representative of the entire highly constrained
parameter space of No-Scale ${\cal F}$-$SU(5)$.  We emphasize that this universal rescaling is not generically available
in models of the Constrained MSSM (CMSSM), {\it i.e.}~mSUGRA, variety.

The supersymmetry breaking parameters for this point slightly differ from previous ${\cal F}$-$SU(5)$ studies~\cite{Li:2010ws,Li:2010mi,Li:2011dw},
insomuch as more precise numerical calculations have been incorporated into our baseline algorithm, with the spectrum also accordingly adjusted.
The masses shift a few GeV from the spectra given in those previous works, but where there are differences, we believe this to be
the more accurate representation.  It should be stated that the branching ratios and decay modes of the spectrum in Table~\ref{tab:masses} of this work and
the spectra in~\cite{Li:2010ws,Li:2010mi,Li:2011dw} are identical, so all related physical properties and signatures studied here
subsequent to the code improvement  will be common to the spectrum of this work as well as the spectra published in previous papers.

The most significant asset of this spectrum for our analysis is the relationship between the stop, gluino, and other squarks.
The light stop and gluino masses are clearly evident in the spectrum of Table~\ref{tab:masses}, as are the heavy squarks.
This distinctive mass pattern of $m_{\widetilde{t}} < m_{\widetilde{g}} < m_{\widetilde{q}}$ is the smoking gun signature, as we shall
shortly argue, and possibly a unique characteristic of only ${\cal F}$-$SU(5)$. To compare the $\cal{F}$-$SU(5)$ model studied
here with more standard Minimal Supersymmetric Standard Model (MSSM) varieties, we have examined the ten ``Snowmass Points and Slopes'' (SPS)
benchmark points~\cite{Allanach:2002nj}.  In an effort to choose a suitable sample, we have limited consideration to those few points featuring
spectra heavy enough to have thus far escaped exclusion by the initial LHC data, yet sufficiently light for potential production at LHC
in the first running year.  We select the mSUGRA point SPS SP3 for our analysis, having also directly verified that SPS SP1B demonstrates
a parallel phenomenology.

We emphasize that although internally stable across our model, the spectrum is substantively different from
each of the ten Snowmass benchmark scenarios.  Of particular note, we find that none of the ten
standardized SPS benchmarks support the $m_{\widetilde{t}} < m_{\widetilde{g}} < m_{\widetilde{q}}$ mass pattern.
This critical ingredient is indicative of how unique the ${\cal F}$-$SU(5)$ signal could be, and just as importantly, how potentially
inadequate the previous LHC SUSY studies could prove with respect to discovery of that signal.

The mechanism of this distinctive signature may be traced to the fact that the one-loop $\beta$-function for the
$SU(3)_C$ gauge symmetry is zero due to the extra vector-like particle contributions~\cite{Jiang:2006hf}.
The effect on the colored gaugino is direct in the running down from the high energy boundary,
leading to the relation $M_3/M_{1/2} \simeq \alpha_3(M_{\rm Z})/\alpha_3(M_{32}) \simeq \mathcal{O}\,(1)$.
Consequently, the low energy gluino mass is lighter than that of all the squarks except for the light
stop.  Since the gluino mass is around $560$~GeV, we anticipate that No-Scale $\cal{F}$-$SU(5)$
may be tested definitively during the early LHC run.  The vector-like fields $M_V$ postulated
in our model have masses around $1000$~GeV, which may be too heavy for immediate direct production.
However, discovery of the distinctively light gluino is in and of itself a highly suggestive
indicator for the role of the vector-like fields, and indeed for the entire stringy origin of No-Scale $\cal{F}$-$SU(5)$.

We would be remiss to overlook some comment on the light stau mass in Table~\ref{tab:masses}, and its implications.
The provided spectrum does indeed exceed the LEP constraints on the lightest neutralino $\widetilde{\chi}_{1}^{0}$ and
lightest stau $\widetilde{\tau}_{1}$~\cite{LEP}, albeit quite narrowly in the latter case.  Of course, our spectrum carries
intrinsic error and corresponding bounds of confidence, which may overlap the experimental bounds.  Moreover, there remains
some very limited freedom to slightly elevate the stau mass, in association with a rescaling of the vector-like fields, such
that proximity to the allowed boundary does not concern us.  On the contrary, we are tantalized by the prospect of
a possible near term discovery of the light stau at the LHC.  Its presence could be reconstructed, for instance, from the
dominant ${\cal F}$-$SU(5)$ process $\widetilde{g} \rightarrow \widetilde{t}_{1} \overline{t} \rightarrow b \overline{t}
\widetilde{\chi}_1^{+} \rightarrow W^{-} b \overline{b} \widetilde{\tau}_{1}^{+} \nu_{\tau}
\rightarrow W^{-} b \overline{b} \tau^{+} \nu_{\tau} \widetilde{\chi}_1^{0}$.
The inference of the short-lived stau from the ${\cal F}$-$SU(5)$ SUSY breaking scenario via tau production assumes
fruition of the expectations for a much improved tau detection efficiency at LHC.

%%%%%%%%%%%%%%%%%%%%%%%%%%%%%%%%%%%%%%%%%%%%%%%%%%%%%%%%%%%%%%%%%%%%%%%%%%%%

\section{A Tool For Selection Cuts\label{sct:cuts}}

Before proceeding to document the ultra-high jet signal of $\cal{F}$-$SU(5)$
in detail, we pause here to interject a description of the user 
adjustable functionality which is accessible within our
selection cut and statistics processing script {\tt CutLHCO}~\cite{cutlhco},
summarizing also the generic output content and form.
The Perl source code has been released into the public domain at the cited web address
to facilitate the rapid prototyping and analysis of alternate selection cut criteria
against Monte Carlo collider and detector simulation data.

In choosing the qualitative character and quantitative tuning of the baseline 
selection cuts to be employed in the reduction of signal backgrounds, we have primarily followed the lead of the
initial multi-jet search strategies favored by the CMS collaboration~\cite{Khachatryan:2011tk,PAS-SUS-09-001}.
Default values for all parameters are given by the ``CMS Style'' column of Table~\ref{tab:cuts}.
Of course, our treatment is of equally broad applicability to the sister ATLAS detector collaboration.
We shall however employ a shorthand language in this work which broadly equates the ``CMS'' selection criteria with
any set of cuts designed and optimized for the resolution of intermediate jet signals from the SM background.
The ``Ultra Jet'' column presents our suggestion of a modified set of selection criteria which are designed
to alternately highlight the presence of an ultra-high jet multiplicity signal, as elaborated in Section~\ref{sct:retune}.

\begin{table}[htbp]
	\centering
	\caption{We list the full parameter specification of our emulation of the default CMS SUSY search
	strategy, along with our suggested alternative for the isolation of ultra-high jet multiplicity events.}
		\begin{tabular}{|c|c|c|} \hline
~~Cut Name~~ & ~~CMS Style~~ & ~~Ultra Jet~~ \\ \hline
{\tt CUT\_FEM} & 0.9 & 10 \\ \hline
{\tt CUT\_PRC} & 3 & 3 \\ \hline
{\tt CUT\_PTS} & 30 & 10 -- 20 \\ \hline
{\tt CUT\_PTC} & 50 & 10 -- 20 \\ \hline
{\tt CUT\_JET} & 3 & $\ge 9$ \\ \hline
{\tt CUT\_PTL} & 100 & 100 \\ \hline
{\tt CUT\_HTC} & 350 & 350 \\ \hline
{\tt CUT\_MET} & 150 & 150 \\ \hline
{\tt CUT\_PRL} & 2 & 2 \\ \hline
{\tt CUT\_ATC} & 0.55 & 0 \\ \hline
{\tt CUT\_RTC} & 1.25 & 10 \\ \hline
{\tt CUT\_PHI} & 0 & 0 \\ \hline
{\tt CUT\_PHC} & 25 & 25 \\ \hline
{\tt CUT\_EMC} & 10 & 10 \\ \hline
		\end{tabular}
\label{tab:cuts}
\end{table}

There are several parameters which may exclude individual event fragments which {\tt PGS4} has classified as jets
from inclusion in a more rigorous internal jet definition.  The input {\tt CUT\_FEM} specifies the maximal electromagnetic fraction
which a jet may possess, calculated as ${(1+ {\rm had}/{\rm em})}^{-1}$ from the hadronic to electromagnetic
calorimeter deposition ratio provided by {\tt PGS4}.  The factor {\tt CUT\_PRC} specifies the maximum (absolute value) pseudo rapidity
$\eta \equiv - \ln \tan(\theta/2)$ which a jet may possess.  The zenith angle $\theta$ is measured from the instantaneous
direction of travel of the counterclockwise beam element, such that forward (or backward) scattering corresponds to
$\eta$ equals plus (or minus) infinity, while $\eta = 0$ is a purely transverse scattering event.  The detector
geometry prevents calorimeter coverage in close angular proximity to the beamline, typically leading to a restriction
on $\eta$ values above about three.  The value of {\tt CUT\_PTS}, in GeV, specifies a soft cut on jet transverse momentum; Passing
this cut allows inclusion in the denominator of the statistic $R(H^{\rm miss}_{\rm T})$, to be described shortly.  The
input {\tt CUT\_PTC} is the hard lower bound on transverse momentum for full classification as a surviving jet.

We must also describe statistics and cuts which apply globally, to the event as a whole.  First, if a jet passes the hard {\tt CUT\_PTC}
cut, but fails either {\tt CUT\_FEM} or {\tt CUT\_PRC}, the event is discontinued.  The value of {\tt CUT\_PRL} specifies the maximum pseudorapidity
of the leading jet, independently of the prior general jet definition.  There is also a simple parameter, {\tt CUT\_JET}, which
specifies the minimum number of surviving jets which an event must have in order to proceed in the analysis; There is a
hard lower bound of two.  Similarly straightforward are {\tt CUT\_PTL} and {\tt CUT\_HTC}, which specify, respectively, the minimum
transverse momentum magnitude for each of the two leading jets, and the minimum net scalar sum on transverse momentum
$H_{\rm T} \equiv \sum_{\rm jets} \left| {\vec{p}}_{\rm T} \right|$ for all jets, both in GeV.
We designate $\eta^*$ as the pseudo-rapidity of the hardest jet.

The input {\tt CUT\_MET} specifies
the minimum ``missing transverse energy'' of the event $H_{\rm T}^{\rm miss}$, again in GeV.  This quantity is defined as the
magnitude of the uncanceled portion of the vector sum over transverse momentum, where $\phi$ is the jet azimuthal angle.
\begin{equation}
H_{\rm T}^{\rm miss} \equiv \sqrt{{\left( \sum_{\rm jets} p_{\rm T} \cos \phi \right)}^2 + {\left( \sum_{\rm jets} p_{\rm T} \sin \phi \right)}^2}
\label{EQ:HTM}
\end{equation}
To survive the {\tt CUT\_MET} cut, an event must pass for $H_{\rm T}^{\rm miss}$ as calculated for the hard jets alone, as
calculated with inclusion of the classified soft jets, and also as natively reported by {\tt PGS4} itself.
A statistic designated as the ``effective mass'' $M_{\rm eff}$ is also calculated for each event.  It is quite similar in
structure to $H_{\rm T}$, except that the scalar sum includes all beam fragments, not only those designated as jets,
and in particular, those reconstructed by {\tt PGS4} as carriers of missing transverse energy, {\it \`a la} Eq.~(\ref{EQ:HTM}).

In some cases, it may be that the appearance of missing energy arises simply because softer jets which might have
in actuality helped to rebalance the $H_{\rm T}^{\rm miss}$ accounting were erroneously discarded.  Therefore, the parameter
{\tt CUT\_RTC} limits the maximum ratio $R(H^{\rm miss}_{\rm T})$ by which the calculation of missing transverse energy for hard jets
may exceed the corresponding value when softer jets are reincorporated.  Another most useful and interesting statistic, generally
denoted as $\alpha_{\rm T}$, has been devised to help distinguish actual missing transverse energy from detector mismeasurements.
\begin{equation}
\alpha_{\rm T} \equiv \frac{1}{2} \left\{ \frac{ 1- \left( \Delta H_{\rm T}^{\rm MIN} / H_{\rm T} \right)}{1 - {\left( H_{\rm T}^{\rm miss} / H_{\rm T} \right)}^2} \right\}
\label{EQ:ALPHAT}
\end{equation}
In the prior, $\Delta H_{\rm T}$ is the (positive) difference in the net scalar transverse momentum between two arbitrarily
partitioned groupings of the surviving jets.  All such possible combinations of pseudo jets are considered, and the minimal value of
$\Delta H_{\rm T}$ is employed in Eq.~(\ref{EQ:ALPHAT}).  If there is no mismeasurement or true missing energy, the value
of $\alpha_{\rm T}$ will just be $1/2$.  For energy mismeasurements of otherwise anti-parallel (pseudo) jet pairs,
subtraction of the nonvanishing scalar difference $\Delta H_{\rm T}$ will tend to drive $\alpha_{\rm T}$ below the midline.  Genuine
missing energy, as manifest in the departure from (pseudo) jet anti-parallelism, will imbalance the vector sum within the factor
$H_{\rm T}^{\rm miss}$ of the denominator, tending to create a contrasting elevation in $\alpha_{\rm T}$ above one-half.  The cut {\tt CUT\_ATC}
places a lower bound on the $\alpha_{\rm T}$ ratio.

A third statistic of significant interest for the isolation of mismeasured
jets is the ``biased'' $\Delta \phi^*$ value, which effectively tests whether the energy balance might be restored by
a jet rescaling.  For each surviving jet in turn, $\Delta \phi^i$ registers the absolute azimuthal angle in the range $(0,\pi)$ which
separates the transverse momentum vector of the $i^{th}$ jet from the negation of the directional imbalance which arises by omitting that
jet from the vector transverse momentum sum.  The minimal such value, denoted with the index ``$*$'' is the one reported.
If a single jet mismeasurement is indeed dominantly responsible for a false missing energy signal, then $\Delta \phi^*$ should register close to zero.
The parameter {\tt CUT\_PHI} will discard events whose minimum $\Delta \phi^*$ is below the specified value. 
Finally, {\tt CUT\_PHC} and {\tt CUT\_EMC} cut events with respectively photons or light leptons (electron, muon) possessing a transverse
momentum above the specification, in GeV.

To support the extraction of actionable information, {\tt CutLHCO} generates
a summary report of surviving events per total jet count, and likewise also
for $b$-tagged jets, total lepton count, and $\tau$-specific leptonic counts in
events with at least two $b$-tagged jets.  The percentage of activity for each of the
Table~\ref{tab:cuts} cuts is documented, along with its percentage as the uniquely
enforced cut on a given event.  In addition, a sorted per-event manifest tabulates
the counts of all jets, $b$-tagged jets, all isolated leptons and
$\tau$-flavored leptons with $\ge 2~b$-tagged jets,
plus the computed statistics for $M_{\rm eff}$, $H_{\rm T}$, $H_{\rm T}^{\rm miss}$, $\eta^*$,
$\alpha_{\rm T}$, $R(H^{\rm miss}_{\rm T})$ and $\Delta \phi^*$.

%%%%%%%%%%%%%%%%%%%%%%%%%%%%%%%%%%%%%%%%%%%%%%%%%%%%%%%%%%%%%%%%%%%%%%%%%%%%%%%%%%%%%%%%

\section{Retuning for Ultra-High Jets\label{sct:retune}}

The bulk of prior detector level MSSM studies have been focused on signals from a low
to intermediate multiplicities of jets, as dictated by the spectra of the mSUGRA, Gauge Mediated Supersymmetry Breaking (GMSB)
and Anomaly Mediated Symmetry Breaking (AMSB) models.
By contrast, the ${\cal F}$-$SU(5)$ with vector-like particles mass pattern of $m_{\widetilde{t}} < m_{\widetilde{g}} < m_{\widetilde{q}}$ produces
events with a high multiplicity of virtual stops, which in turn concludes in events with a very large number of jets through
the dominant chains $\widetilde{g} \rightarrow \widetilde{t}_{1} \overline{t} \rightarrow t \overline{t} \widetilde{\chi}_1^{0}
\rightarrow W^{+}W^{-} b \overline{b} \widetilde{\chi}_1^{0}$ and $\widetilde{g} \rightarrow \widetilde{t}_{1} \overline{t}
\rightarrow b \overline{t} \widetilde{\chi}_1^{+} \rightarrow W^{-} b \overline{b} \widetilde{\tau}_{1}^{+} \nu_{\tau}
\rightarrow W^{-} b \overline{b} \tau^{+} \nu_{\tau} \widetilde{\chi}_1^{0}$, as well as the conjugate processes
$\widetilde{g} \rightarrow \widetilde{\overline{t}}_{1} t \rightarrow t \overline{t} \widetilde{\chi}_1^{0}$ and $\widetilde{g}
\rightarrow \widetilde{\overline{t}}_{1} t \rightarrow \overline{b} t \widetilde{\chi}_1^{-}$, where the $W$ bosons will produce
mostly hadronic jets and some leptons. Additionally, the heavy squarks will produce gluinos by means of $\widetilde{q} \rightarrow q \widetilde{g}$.

\begin{figure}[htbp]
        \centering
        \includegraphics[width=0.45\textwidth]{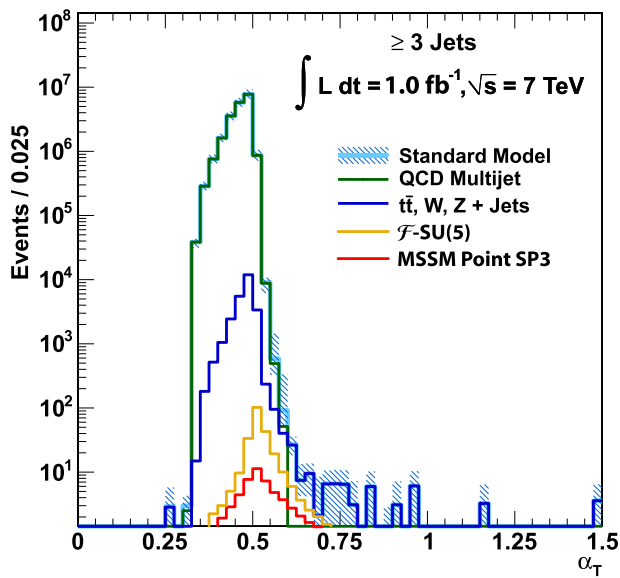}
        \caption{The figure depicts implementation of the CMS style cuts of~\cite{Khachatryan:2011tk,PAS-SUS-09-001}
	on the ${\cal F}$-$SU(5)$ and SPS SP3 Monte Carlo simulations.
	Our histograms, binned at intervals of $0.025$ in $\alpha_{\rm T}$, are superimposed onto the leading SM backgrounds previously published by
	the CMS collaboration~\cite{Khachatryan:2011tk}, with the vertical axis rescaled appropriately for $1~{\rm fb}^{-1}$ of luminosity.}
\label{fig:FSU5_SP3_CMS}
\end{figure}

Systematically employing the procedure detailed in Section~\ref{sct:ccc}, we have modeled the detector environment of the early operational
phase of the LHC, generating Monte Carlo events at a center of mass energy ${\sqrt s}=7$ TeV for the No-Scale $\cal{F}$-$SU(5)$ and mSUGRA SPS SP3 benchmark
points of Section~\ref{sct:spectrum}.  The accumulated histogram counts for all events with three or more jets,
post-processed under the CMS style cuts~\cite{Khachatryan:2011tk,PAS-SUS-09-001}
of Table~\ref{tab:cuts} from Section~\ref{sct:cuts} (excepting the cut on the plotting variable $\alpha_{\rm T}$),
are superimposed in Fig.~(\ref{fig:FSU5_SP3_CMS}) onto the corresponding SM backgrounds
borrowed from Ref.~\cite{Khachatryan:2011tk}, with the vertical axis rescaled for $1~{\rm fb}^{-1}$ of luminosity.
The devastating consequence of imposing the baseline CMS style cuts onto ${\cal F}$-$SU(5)$ is
unmistakably revealed, as the signature is entirely concealed behind the dominant SM contribution.
The Snowmass benchmark is likewise even more strongly suppressed, but we shall see by contrast that it
has no better hope for redemption with respect to the ultra-high jet multiplicity counts.

The propensity for ultra-high jet events was clear from the outset of our collider simulation efforts.  We
recognized that a separation of jet counts bin-by-bin demonstrated a clear signal for No-Scale $\cal{F}$-$SU(5)$ in
the high jet multiplicities $(6,7,8,\ldots)$, whereas the clustering of all jets into a unified statistic shrouded the signal
behind an barrage of surviving intermediate count $(3,4,5)$ events from the background, as effectively demonstrated
by Fig.~(\ref{fig:FSU5_SP3_CMS}).  We purposed then to pursue a simple strategy for retuning our cuts in a manner which would even more
strongly emphasize the high and ultra-high $(9,10,11,12,\ldots)$ jet content.  Since the ultra-high jet regime is greatly
suppressed in the SM backgrounds, we were able to relax certain of the harsh cuts which are very effective for separating
out the MSSM in intermediate jet searches, but which simultaneously exert a costly attrition against our signal.

Table~\ref{tab:cuts} further compares our attempt to mimic the CMS style cuts, which are optimized for an intermediate jet count
search, against our proposal for an ultra-high jet search strategy.  Specifically, we effectively disable the cuts on
electromagnetic fraction, $\alpha_{\rm T}$ (as in Eq.~\ref{EQ:ALPHAT}), and the missing energy ratio $R(H_{\rm T}^{\rm miss})$ of hard to soft jets.
In addition, and most significantly, we reduce the threshold on missing transverse momentum per jet to either $10$~GeV or $20$~GeV for both the hard
(previously $50$~GeV) and soft (previously $30$~GeV) jet categories, although the two leading jets are still required to carry $100$~GeV of transverse
momentum each, and the limit on net transverse momentum is unchanged.

\begin{figure*}[htbp]
        \centering
        \includegraphics[width=1.00\textwidth]{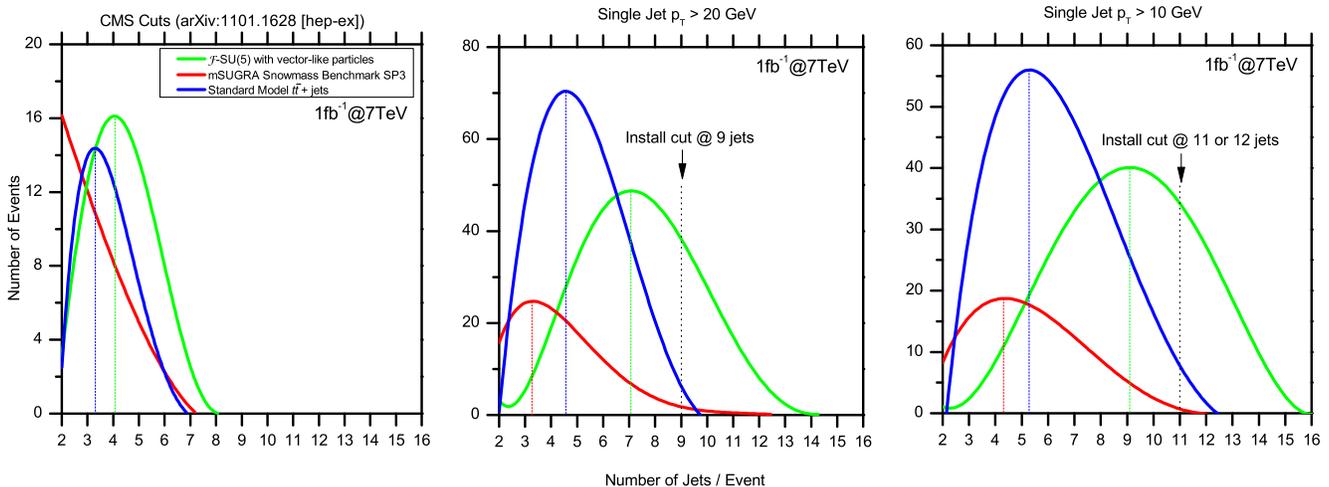}
        \caption{Distribution of events per number of jets. A polynomial fit has been interpolated over the histogram data.}
\label{fig:jet_comp}
\end{figure*}

In Fig.~(\ref{fig:jet_comp}) we plot the number of jets per event versus the number of events for a triplet of distinct scenarios.
To suppress histogram noise and emphasize the peak in jet multiplicity, we interpolate a polynomial fit over the data points.
Within each of the three panes, the No-Scale $\cal{F}$-$SU(5)$ benchmark, the SPS SP3 mSUGRA benchmark, and the leading SM
$t \overline{t} + {\rm jets}$ background are each represented.  The first pane displays a comparison of the number
of jets when employing the canonical CMS style cuts of~\cite{Khachatryan:2011tk,PAS-SUS-09-001}, which clearly downgrade all
of the ${\cal F}$-$SU(5)$ ultra-high jet multiplicity events, converting processes which feature 9 or more distinct jets into
events with effectively far fewer.  The latter two panes represent the effort to retain this essential signal information via
alternative selection cuts, shifting to a minimum $p_{\rm T}$ per jet of $20$ or $10$~GeV, respectively.  The final scenario
is perhaps overly close to the onset of severe jet fragmentation, and our greater comfort is with the more conservative
$20$~GeV selection.

It is clear graphically that this alternate prescription accomplishes the dual goals of elevating the
peak numerical jet acquisition per bin, and shifting the location of the peak to a larger count of jets.
The figures as plotted further allow us to gauge an appropriate selection cut for the number of jets to maximize our signal to background ratio,
while assessing the impact of the CMS style selection cuts upon the model studied in this work.
We see that both the 10 GeV and 20 GeV jet $p_{\rm T}$ cuts preserve the high number of jets,
permitting an obvious choice for location of the cut on the minimum number of jets.

For jet $p_{\rm T} >$ 20 GeV, the signal to background ratio is maximized for greater than 9 jets. Retaining only events with greater
than 10 jets is satisfactory as well, however the small gain in background suppression may not be worth the more significant
reduction in net events.  Examining the situation for jets of $p_{\rm T} >$ 10 GeV, we reach similar conclusions, though this time the cut
can be placed at 11 or 12 jets. We thus adopt four distinct revised cuts of single jet $p_{\rm T} >$ 20 GeV and total number of
jets greater than 9 and likewise greater than 10, and single jet $p_{\rm T} >$ 10 GeV and total number of jets greater than 11 and similarly greater than 12.
It is apparent that the cuts optimized for intermediate jet multiplicities are quite debilitating in comparison, and place in jeopardy any potential
high jet signal which might then effectively hide in plain sight.

In Table~\ref{tab:cutrate}, we present a detailed breakdown of the percentage of activity of each of the cut parameters outlined 
in Table~\ref{tab:cuts}, for both the baseline CMS style criteria, and the ultra-high jet ($p_{\rm T} > 20$, ${\rm jets} \ge 9$) 
scenario, which will become our principal operational default.  Within each primary subdivision, and for each of the ${\cal F}$-$SU(5)$,
SPS~SP3 and $t\overline{t}+{\rm jets}$ data sets, we report both the total rate of application for each cut, and the percentage of
events for which each cut represents the sole reason of exclusion.  For both of the post-SM models, the strongest single cut in
the CMS styled event processing is that on $\alpha_{\rm T}$;  Under the ultra-high styled processing the dominant role is, not
surprisingly, played by the jet count limit itself.  Both scenarios are extraordinarily effective, almost to totality, against
the SM subset which we have modeled, although we should remark that for the lower jet search strategies, the $t\overline{t}+{\rm jets}$
processes do not sufficiently represent the full SM.

\begin{table*}[htbp]
  \small
	\centering
	\caption{Percentage of activity of each cutting parameter for the CMS style and baseline ($p_{\rm T} > 20$, ${\rm jets} \ge 9$) ultra-high
		jet search strategies, for each of the ${\cal F}$-$SU(5)$, SPS~SP3, $t\overline{t}+{\rm jets}$ process simulations.  Each column
		is split to provide both the overall activity, and the percentage of events for which the given cut is a unique discriminant.}
		\begin{tabular}{c|r|r||r|r||r|r||r|r||r|r||r|r|} \cline{2-13}
\multicolumn{1}{c}{} & \multicolumn{6}{|c||}{\rm CMS Style Cut Percentages} & \multicolumn{6}{|c|}{Ultra-High Jet Cut Percentages} \\ \hline 
\multicolumn{1}{|c||}{~{\rm Cut~Name}~} & \multicolumn{2}{|c||}{~~${\cal F}$-$SU(5)$~~} & \multicolumn{2}{|c||}{~~SPS~SP3~~} & \multicolumn{2}{|c||}{~~$t\overline{t}+{\rm jets}$~~} &
           \multicolumn{2}{|c||}{~~${\cal F}$-$SU(5)$~~} & \multicolumn{2}{|c||}{~~SPS~SP3~~} & \multicolumn{2}{|c|}{~~$t\overline{t}+{\rm jets}$~~} \\ \hline \hline 
\multicolumn{1}{|c||}{$~{\tt CUT\_(FEM|PRC)}~$} & $~1.07~$ & $~0.07~$ & $~1.83~$ & $~0.18~$ & $~1.02~$ & $~0.00~$ & $~0.11~$ & $~0.00~$ & $~0.06~$ & $~0.00~$ & $~0.15~$ & $~0.00~$ \\ \hline
\multicolumn{1}{|c||}{$~{\tt CUT\_JET}~$} & $~63.39~$ & $~0.15~$ & $~55.71~$ & $~5.08~$ & $~50.91~$ & $~0.00~$ & $~88.31~$ & $~9.57~$ & $~98.28~$ & $~34.68~$ & $~98.58~$ & $~0.37~$ \\ \hline
\multicolumn{1}{|c||}{$~{\tt CUT\_PTL}~$} & $~66.59~$ & $~0.37~$ & $~36.64~$ & $~0.36~$ & $~78.22~$ & $~0.02~$ & $~66.52~$ & $~0.26~$ & $~36.35~$ & $~0.03~$ & $~78.11~$ & $~0.00~$ \\ \hline
\multicolumn{1}{|c||}{$~{\tt CUT\_HTC}~$} & $~64.42~$ & $~0.09~$ & $~31.65~$ & $~0.05~$ & $~80.94~$ & $~0.02~$ & $~62.41~$ & $~0.00~$ & $~30.21~$ & $~0.00~$ & $~68.97~$ & $~0.00~$ \\ \hline
\multicolumn{1}{|c||}{$~{\tt CUT\_MET}~$} & $~77.41~$ & $~0.03~$ & $~35.63~$ & $~0.00~$ & $~98.12~$ & $~0.01~$ & $~75.94~$ & $~3.73~$ & $~34.84~$ & $~0.16~$ & $~97.78~$ & $~0.68~$ \\ \hline
\multicolumn{1}{|c||}{$~{\tt CUT\_PRL}~$} & $~51.24~$ & $~0.05~$ & $~20.37~$ & $~0.07~$ & $~10.09~$ & $~0.00~$ & $~39.07~$ & $~0.05~$ & $~14.96~$ & $~0.01~$ & $~6.99~$ & $~0.00~$ \\ \hline
\multicolumn{1}{|c||}{$~{\tt CUT\_ATC}~$} & $~92.64~$ & $~9.39~$ & $~73.22~$ & $~13.45~$ & $~90.17~$ & $~0.22~$ & $~0.00~$ & $~0.00~$ & $~0.00~$ & $~0.00~$ & $~0.00~$ & $~0.00~$ \\ \hline
\multicolumn{1}{|c||}{$~{\tt CUT\_RTC}~$} & $~5.27~$ & $~0.06~$ & $~1.78~$ & $~0.01~$ & $~32.92~$ & $~0.00~$ & $~0.00~$ & $~0.00~$ & $~0.00~$ & $~0.00~$ & $~0.00~$ & $~0.00~$ \\ \hline
\multicolumn{1}{|c||}{$~{\tt CUT\_PHI}~$} & $~0.00~$ & $~0.00~$ & $~0.00~$ & $~0.00~$ & $~0.00~$ & $~0.00~$ & $~0.00~$ & $~0.00~$ & $~0.00~$ & $~0.00~$ & $~0.00~$ & $~0.00~$ \\ \hline
\multicolumn{1}{|c||}{$~{\tt CUT\_PHC}~$} & $~1.26~$ & $~0.04~$ & $~2.22~$ & $~0.20~$ & $~1.43~$ & $~0.00~$ & $~1.26~$ & $~0.07~$ & $~2.22~$ & $~0.02~$ & $~1.43~$ & $~0.00~$ \\ \hline
\multicolumn{1}{|c||}{$~{\tt CUT\_EMC}~$} & $~22.28~$ & $~1.14~$ & $~42.32~$ & $~5.02~$ & $~32.45~$ & $~0.07~$ & $~22.28~$ & $~1.20~$ & $~42.32~$ & $~0.42~$ & $~32.45~$ & $~0.00~$ \\ \hline \hline
\multicolumn{1}{|c||}{~{\rm Net~Efficiency}~} & \multicolumn{2}{|c||}{97.49} & \multicolumn{2}{|c||}{90.98} & \multicolumn{2}{|c||}{99.96} &
           \multicolumn{2}{|c||}{95.60} & \multicolumn{2}{|c||}{99.06} & \multicolumn{2}{|c|}{99.99} \\ \hline
		\end{tabular}
		\label{tab:cutrate}
\end{table*}

Considering the large number of hadronic jets which are required by our optimized ultra-high jet signatures, there is little intrusion from SM 
background processes after post-processing cuts. We have examined the background processes studied in~\cite{Baer:2010tk,Altunkaynak:2010we}
and assessed the relevance of each to our model in the initial LHC run. Our conclusion is that only the
$t \overline{t} + {\rm jets}$ process possesses the requisite minimum cross-section and multiplicity of final
state jet production to compete with the ${\cal F}$-$SU(5)$ signal. Processes with a larger number of top quarks can also
generate events with a large number of jets, however, the cross-sections are sufficiently suppressed to be negligible, bearing in mind
the large number of ultra-high jet events which our model will generate.  The same is true for those more complicated background processes involving
combinations of top quarks, jets, and one or more vector bosons.

Furthermore, we neglect the pure QCD $(2,3,4)$ jet events, one or more vector boson events, and all $b \overline{b}$ processes,
since none of these can sufficiently produce events with 9 or more jets after post-processing cuts have been applied.
The number of events for these will be quite large, though practically all of the jets from detector effects beyond the
initial hadronization are ultimately discarded.  We intend, again, to more fully address any such lower order
backgrounds which are of some residual relevancy in subsequent publications targeting the higher energy,
larger luminosity, latter operational phases of the LHC.

Certainly it is true that the large count of softer jets which we have considered here do themselves represent a significant fragmentation from
the hard jet showering.  Nevertheless, the basic intuition that fewer hard jets in the early parton level diagrams will
yield a correspondingly smaller count of final state soft jets is well confirmed by the Monte Carlo, and we observe
not only a readily detectable signal for No-Scale $\cal{F}$-$SU(5)$ above the SM background, but also a clear differentiation
between No-Scale $\cal{F}$-$SU(5)$ and a typical competing post-SM scenario.  The unique SUSY mass hierarchy of No-Scale $\cal{F}$-$SU(5)$,
which we have not found replicated by any models of the CMSSM variety leads us to suspect that this conclusion
may be broadly generalized.

We have verified that the cuts proposed in this section remain globally stronger than typical Level~1 triggers.
As such, our suggested selection criteria represent only a modest alternative post-processing phase: a practical
variation upon the theme of the existing search language, requiring no restructuring of the basic data
collection operation, and suggesting no exotic or highly specialized search technology.

%%%%%%%%%%%%%%%%%%%%%%%%%%%%%%%%%%%%%%%%%%%%%%%%%%%%%%%%%%%%%%%%%%%%%%%%%%%%

\section{No-Scale ${\cal F}$-$SU(5)$ Collider Signals\label{sct:signal}}

To more fully assess the discovery potential of our optimized selection cut criteria, and of the No-Scale $\cal{F}$-$SU(5)$
signal in particular, we complete in this section our comparative analysis of the Monte Carlo collider and detector simulation of
No-Scale $\cal{F}$-$SU(5)$ with vector-like particles, the SPS SP3 mSUGRA benchmark, and the leading SM $t \overline{t} + {\rm jets}$
background.  Throughout this study, we have maintained a center of mass energy $\sqrt{s} = 7$~TeV, and have normalized all event counts
to $1~{\rm fb}^{-1}$ of integrated luminosity, in keeping with the net expected LHC data collection yield through the year 2011.
The actual amount of data which we processed in each of the three cases is somewhat larger, corresponding respectively to
$100,000$ events with a total cross-section of $2.125$~pb for $47.1~\rm {fb}^{-1}$ of luminosity ($\cal{F}$-$SU(5)$),
$100,000$ events with a total cross-section of $0.285$~pb for $351~\rm {fb}^{-1}$ of luminosity (SPS SP3),
$\sim 120,000$ events with a total cross-section of $79.8$~pb for $1.50~\rm {fb}^{-1}$ of luminosity ($t \overline{t} + {\rm jets}$),
and has been scaled down by the individually appropriate factor.

\begin{figure}[htbp]
        \centering
        \includegraphics[width=0.45\textwidth]{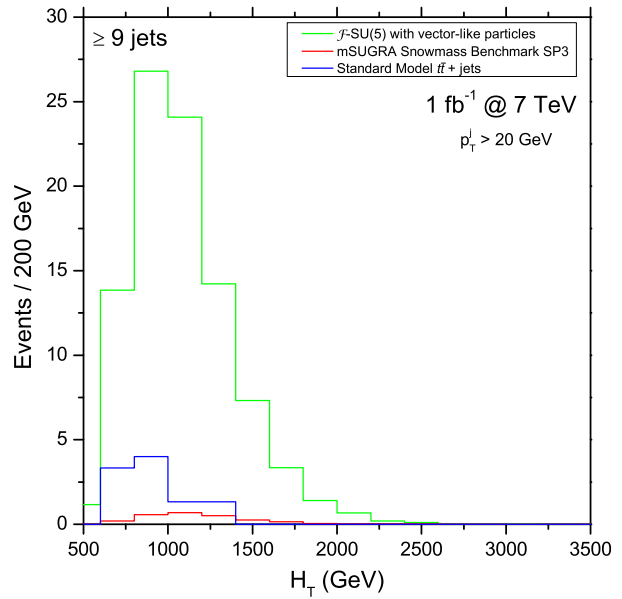}
        \caption{$H_{\rm T}$ for events with $\ge$ 9 jets for 1 ${\rm fb}^{-1}$ and $\sqrt{s}$ = 7~TeV. Minimum $p_{\rm T}$ for a single jet is 20 GeV.}
\label{fig:gte9jets}
\end{figure}

\begin{figure}[htbp]
        \centering
        \includegraphics[width=0.45\textwidth]{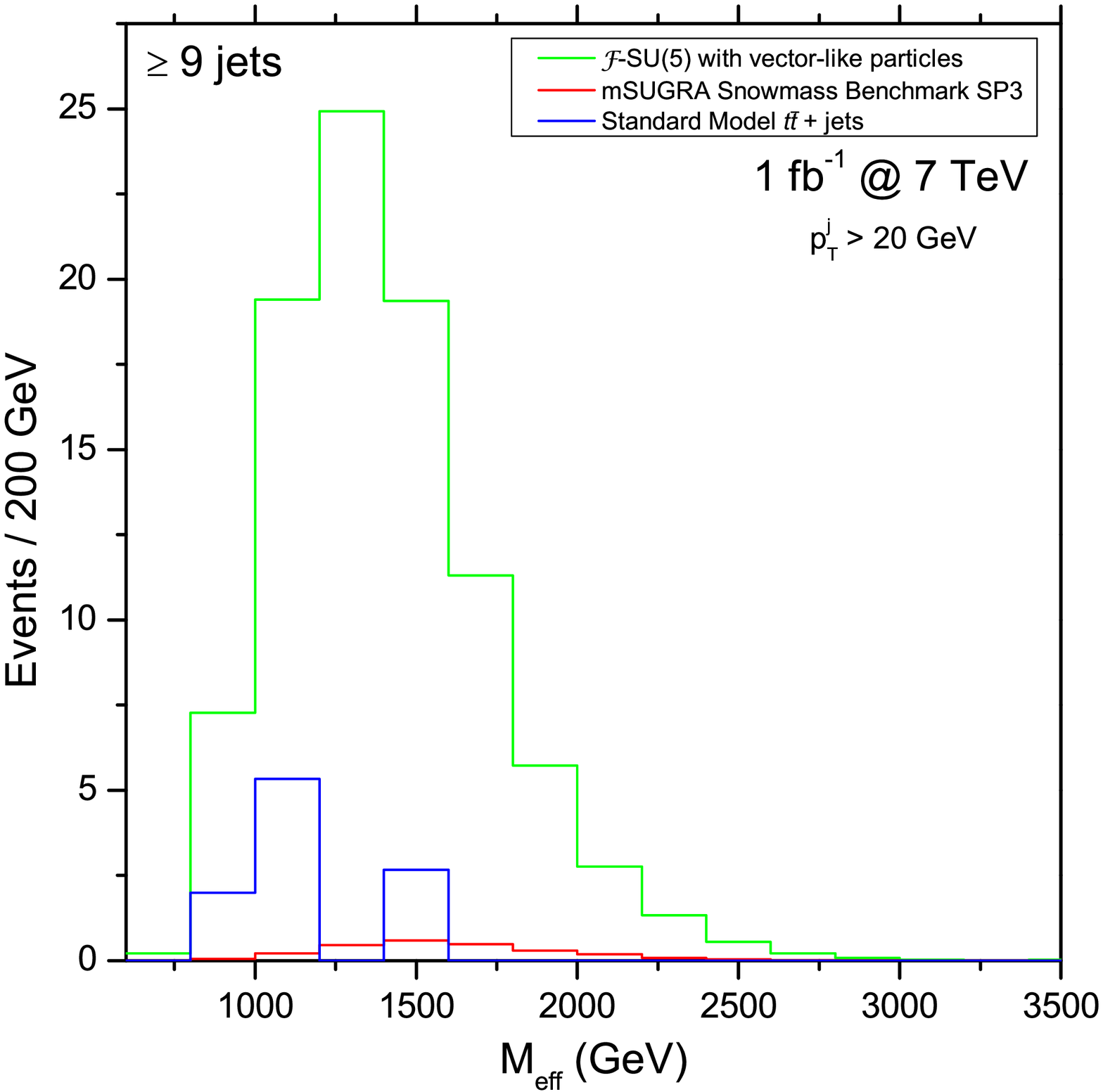}
        \caption{Effective mass for events with $\ge$ 9 jets for 1 ${\rm fb}^{-1}$ and $\sqrt{s}$ = 7~TeV. Minimum $p_{\rm T}$ for a single jet is 20 GeV.}
\label{fig:gte9jets_Meff}
\end{figure}

\begin{figure}[htbp]
        \centering
        \includegraphics[width=0.45\textwidth]{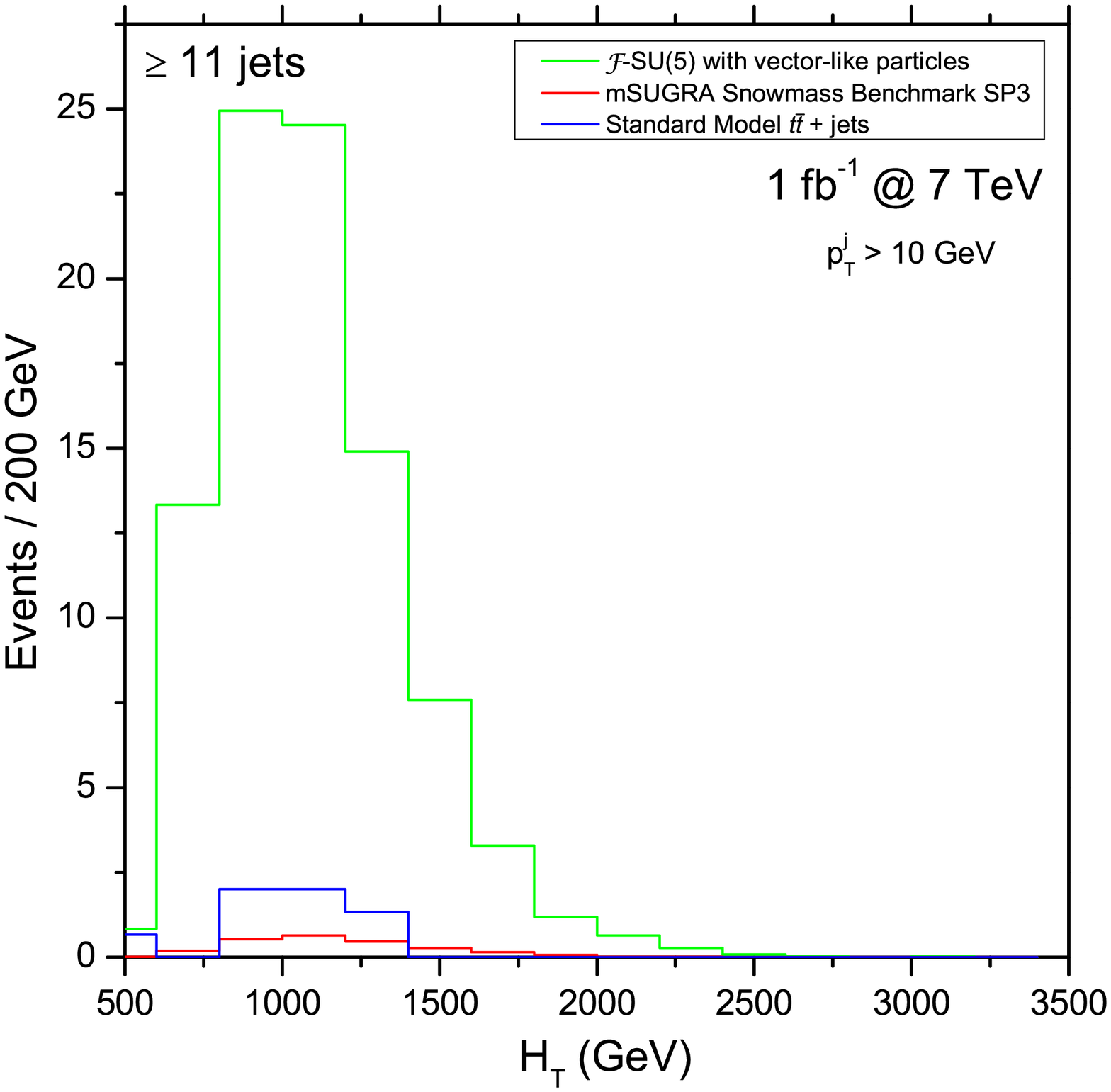}
        \caption{$H_{\rm T}$ for events with $\ge$ 11 jets for 1 ${\rm fb}^{-1}$ and $\sqrt{s}$ = 7~TeV. Minimum $p_{\rm T}$ for a single jet is 10 GeV.}
\label{fig:gte11jets}
\end{figure}

\begin{figure}[htbp]
        \centering
        \includegraphics[width=0.45\textwidth]{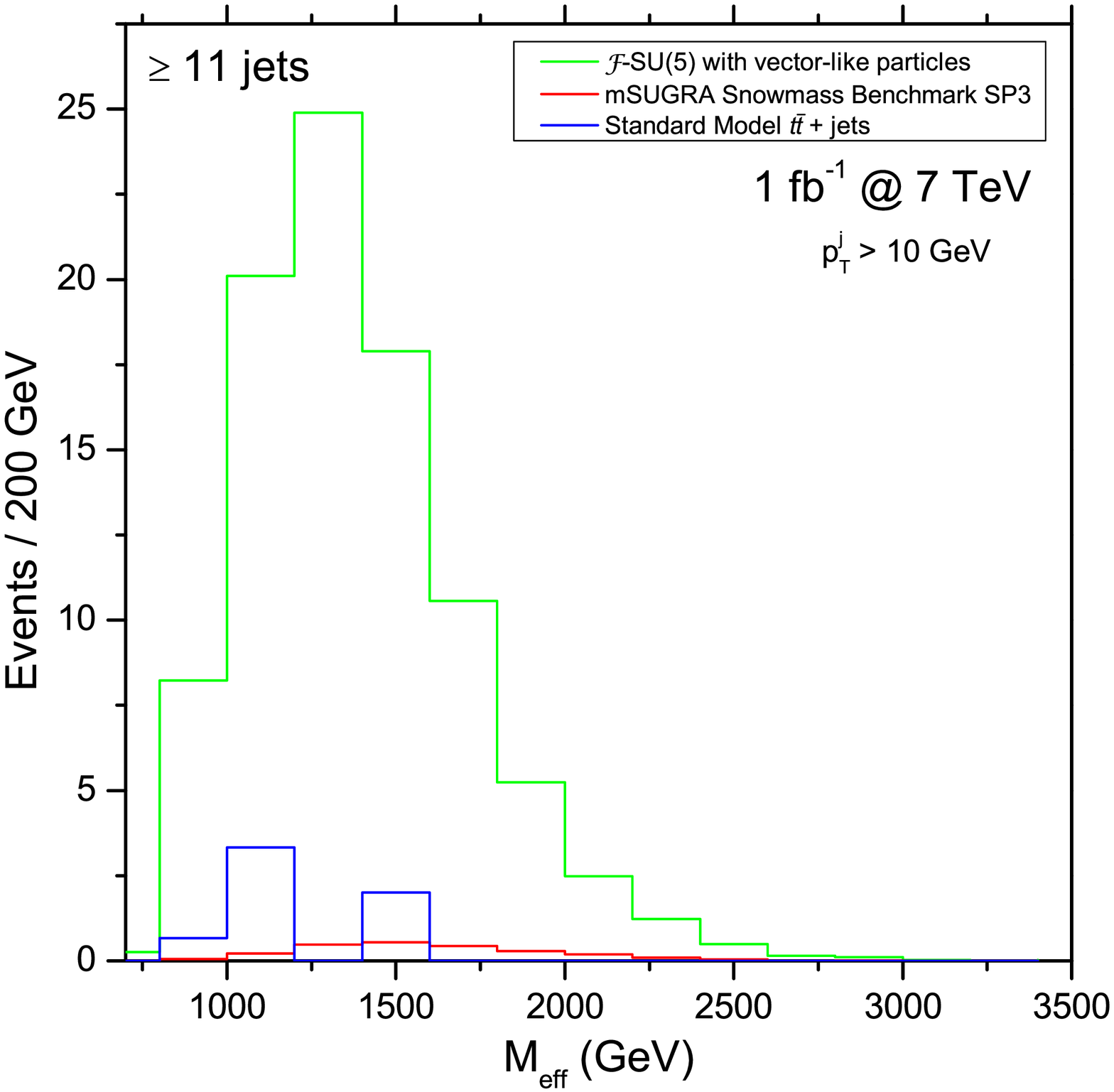}
        \caption{Effective mass for events with $\ge$ 11 jets for 1 ${\rm fb}^{-1}$ and $\sqrt{s}$ = 7~TeV. Minimum $p_{\rm T}$ for a single jet is 10 GeV.}
\label{fig:gte11jets_Meff}
\end{figure}

\begin{figure}[htbp]
        \centering
        \includegraphics[width=0.45\textwidth]{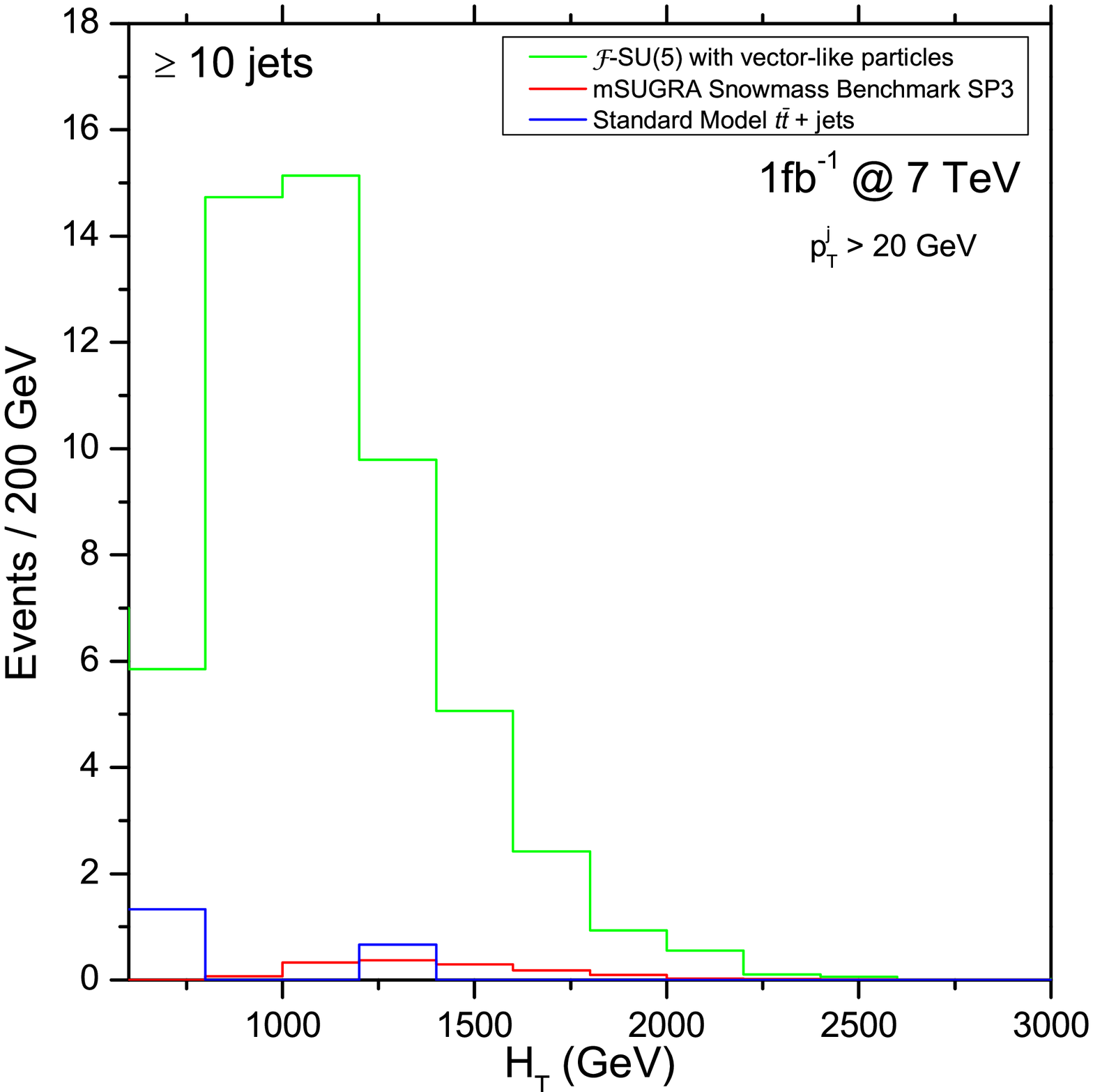}
        \caption{$H_{\rm T}$ for events with $\ge$ 10 jets for 1 ${\rm fb}^{-1}$ and $\sqrt{s}$ = 7~TeV. Minimum $p_{\rm T}$ for a single jet is 20 GeV.}
\label{fig:gte10jets}
\end{figure}

\begin{figure}[htbp]
        \centering
        \includegraphics[width=0.45\textwidth]{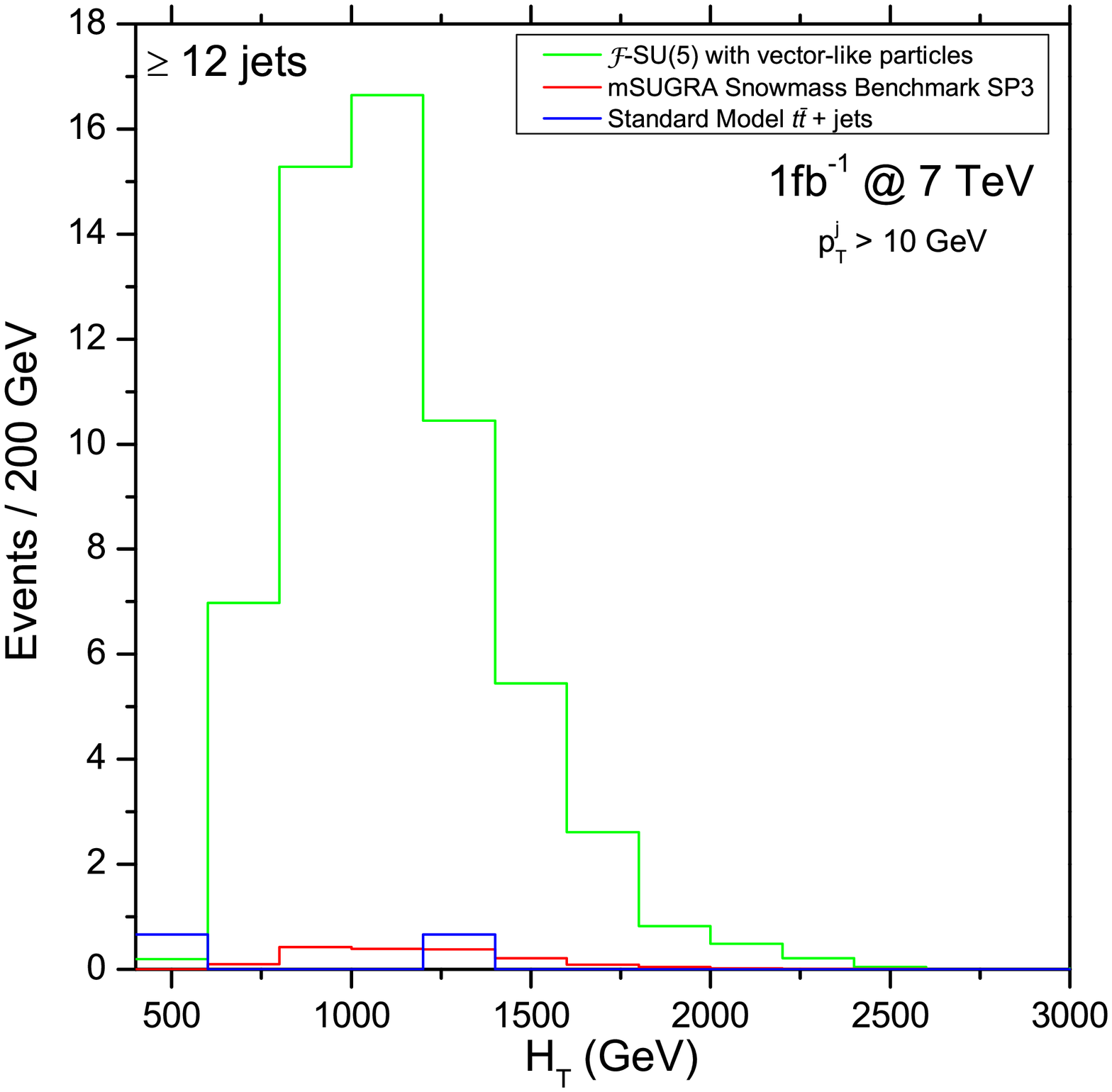}
        \caption{$H_{\rm T}$ for events with $\ge$ 12 jets for 1 ${\rm fb}^{-1}$ and $\sqrt{s}$ = 7~TeV. Minimum $p_{\rm T}$ for a single jet is 10 GeV.}
\label{fig:gte12jets}
\end{figure}

To begin, we plot the number of events per 200 GeV bin size versus $H_{\rm T} \equiv \sum_{\rm jets} \left| {\vec{p}}_{\rm T} \right|$ and
also versus the effective mass $M_{\rm eff} \equiv H_{\rm T} + H_{\rm T}^{\rm miss}$ for the spectrum of
Table~\ref{tab:masses}.  We exhibit $H_{\rm T}$ for all four optimized ultra-high multiplicity jet signatures, though $M_{\rm eff}$ for
only $\ge$ 9 jets and $\ge$ 11 jets signatures, since the similarity of the $M_{\rm eff}$ distribution to the $H_{\rm T}$ distribution is
readily apparent.  Figs.~(\ref{fig:gte9jets} -- \ref{fig:gte12jets}) depict the convincing separation between the ${\cal F}$-$SU(5)$ signal and the SM
$t \overline{t} + {\rm jets}$ background, in addition to a clear distinction from the SPS benchmark point SP3.

We also endeavor to capture in our analysis the large number of $b$-jets generated by ${\cal F}$-$SU(5)$. To more
faithfully emulate the projected CMS and ATLAS performance in observing $b$-jets at around $60\%$ efficiency, we have updated the $b$-tagging
efficiency functions in {\tt PGS4}, maintaining the existing usage of a fifth order polynomial fit, but revising the numerical coefficients as follows:
$b\,(p_{\rm T}) = 0.0883 + 0.0197~p_{\rm T} - 2.4872 \times 10^{-4}~p_{\rm T}^2 +
1.47212 \times 10^{-6}~p_{\rm T}^3 - 4.16484 \times 10^{-9}~p_{\rm T}^4 + 4.41957 \times 10^{-12}~p_{\rm T}^5$ and
$b(\eta) = 1.00885 - 0.04975~\eta + 0.0693~\eta^2 - 0.03611~\eta^3 - 0.02222~\eta^4 + 0.00798~\eta^5$.
To negate the SM $b \overline{b} + {\rm jets}$ and $b \overline{b} b \overline{b}$ processes, we require at least one lepton in the event,
in this case a tau to minimize the background further, along with at least three $b$-jets.

\begin{figure}[htbp]
        \centering
        \includegraphics[width=0.45\textwidth]{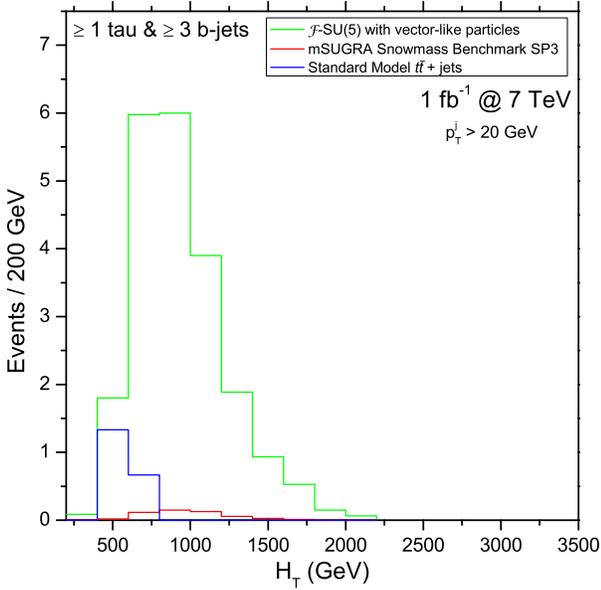}
        \caption{$H_{\rm T}$ for events with $\ge$ 1 tau $\&\ge$ 3 $b$-jets for 1 ${\rm fb}^{-1}$ and $\sqrt{s}$ = 7~TeV. Minimum $p_{\rm T}$ for a single jet is 20 GeV.}
\label{fig:gte1tau_gte3bjet}
\end{figure}

\begin{figure}[htbp]
        \centering
        \includegraphics[width=0.45\textwidth]{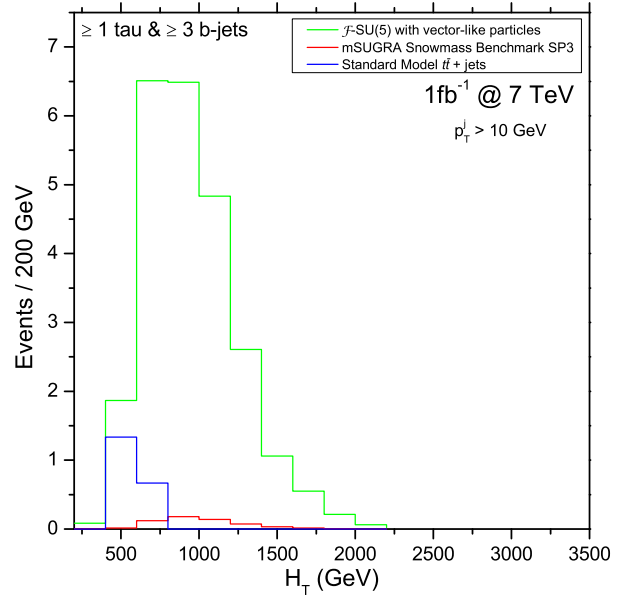}
        \caption{$H_{\rm T}$ for events with $\ge$ 1 tau $\&\ge$ 3 $b$-jets for 1 ${\rm fb}^{-1}$ and $\sqrt{s}$ = 7~TeV. Minimum $p_{\rm T}$ for a single jet is 10 GeV.}
\label{fig:gte1tau_gte3bjet_10GeV}
\end{figure}

The projected counts for these events in ${\cal F}$-$SU(5)$ are smaller than those of the ultra-high multiplicity jet events,
though the signal to background ratio remains quite favorable.  Fig.~(\ref{fig:gte1tau_gte3bjet}) and Fig.~(\ref{fig:gte1tau_gte3bjet_10GeV}),
for the single jet $p_{\rm T}$ $>$ 20 GeV and $p_{\rm T}$ $>$ 10 GeV cases respectively, reveal that the $\ge$ 1 tau and $\ge$ 3 $b$-jets
signature supplements the ultra-high jet signatures very nicely, providing confirmation of the potential for ${\cal F}$-$SU(5)$ signal
discovery. Requiring at least 4 $b$-jets in an event improves the signal to background ratio even further, however, this process
will not generate enough events to be observable in the early LHC run. We thus omit the analysis of $\ge 1$~tau and
$\ge 4$ $b$-jets in this work, though we will plan to explore it in more depth in follow-up studies of future LHC phases.

The number of events for each optimized signature, for each of $\cal{F}$-$SU(5)$, SPS SP3 and $t \overline{t} + {\rm jets}$, are summarized in Table~\ref{tab:counts}.
We also include a standard measure of discovery threshold that compares the number of signal events $S$ to the number of background
events $B$, where $S / \sqrt{B} > 5$ is generally considered to be favorable. Notice that ${\cal F}$-$SU(5)$ comfortably surpasses this requirement,
while the SPS SP3 benchmark is well below the minimum necessary for observability under the umbrella of our post-processing selection cuts and signatures.

\begin{table*}[htbp]
  \small
	\centering
	\caption{Total number of events for $1~{\rm fb}^{-1}$ and 7~TeV for each of our optimized signatures.
	We require $S / \sqrt{B} > 5$, where $S$ is the number of signal events and $B$ the number of background events, $t \overline{t} + {\rm jets}$ in our analysis here.}
		\begin{tabular}{|c||c|c||c|c||c|} \hline
    ${\rm Optimized~Signature}$ & ~~${\cal F}$-$SU(5)$~~ & ~~$\frac{S}{\sqrt{B}}$~~&~~${\rm SPS~SP3}$~~ & ~~$\frac{S}{\sqrt{B}}$~~ & ~~$t\overline{t} + {\rm jets}$~~\\ \hline\hline
    $\ge 9~ {\rm jets}~{\rm and}~p_{\rm T}>20~{\rm GeV}$&$93.2$&$29.5$&$2.7$&$0.85$&$10.0$\\ \hline
    $\ge 10~ {\rm jets}~{\rm and}~p_{\rm T}>20~{\rm GeV}$&$54.7$&$38.7$&$1.4$&$0.99$&$2.0$\\ \hline
    $\ge 11~ {\rm jets}~{\rm and}~p_{\rm T}>10~{\rm GeV}$&$91.6$&$37.4$&$2.4$&$0.98$&$6.0$\\ \hline
    $\ge 12~ {\rm jets}~{\rm and}~p_{\rm T}>10~{\rm GeV}$&$59.2$&$51.9$&$1.7$&$1.5$&$1.3$\\ \hline
    $\ge 1 \tau~\&~\ge 3~{\rm bjets}~{\rm and}~p_{\rm T}>20~{\rm GeV}$&$21.3$&$15.1$&$0.5$&$0.35$&$2.0$\\ \hline
    $\ge 1 \tau~\&~\ge 3~{\rm bjets}~{\rm and}~p_{\rm T}>10~{\rm GeV}$&$24.3$&$17.2$&$0.58$&$0.41$&$2.0$\\ \hline
		\end{tabular}
		\label{tab:counts}
\end{table*}

In Tables~(\ref{tab:rawcms},\ref{tab:rawultrahigh}), for the CMS style and ($p_{\rm T} > 20$, ${\rm jets} \ge 9$)
ultra-high jet search criteria respectively, we provide the raw number of surviving events per distributed count of jets, $b$-tagged jets,
isolated leptons and $\tau$-flavored leptons with $\ge 2~b$-tagged jets.  The tabulated results have been integrally rounded after scaling
to $1~{\rm fb}^{-1}$ of luminosity.  The net count of surviving $\cal{F}$-$SU(5)$ events is demonstrated to be larger under the ultra-high cut scenario,
showcasing a wealth of activity at and above the nine jet threshold, a territory fully excluded under the search strategy optimized for intermediate jet
multiplicities.  The $\cal{F}$-$SU(5)$ model blends inconspicuously into its surroundings under the Table~\ref{tab:rawcms} cuts, while it is prominently
and unmistakably on display in Table~\ref{tab:rawultrahigh}.

\begin{table}[htbp]
  \small
	\centering
	\caption{Distributed integral event counts for $1~{\rm fb}^{-1}$ and $\sqrt{s}=7$~TeV for the CMS style cut criteria.}
		\begin{tabular}{|c||c|c|c|} \hline
\multicolumn{4}{|c|}{\rm Raw CMS Style Event Distribution} \\ \hline
~~Multiplicity~~ & ~~${\cal F}$-$SU(5)$~~ & ~~SPS~SP3~~ & ~~$t\overline{t}+{\rm jets}$~~ \\ \hline
\hline \multicolumn{4}{|c|}{~{\rm Surviving~Events~per~Jet~Count}~} \\ \hline
$3$ & $12$ & $13$ & $13$ \\ \hline
$4$ & $17$ & $8$ & $13$ \\ \hline
$5$ & $14$ & $3$ & $7$ \\ \hline
$6$ & $8$ & $1$ & $2$ \\ \hline
$7$ & $2$ & $0$ & $0$ \\ \hline
$8$ & $1$ & $0$ & $0$ \\ \hline
$~{\rm Net}~$ & $54$ & $25$ & $35$ \\ \hline
\hline \multicolumn{4}{|c|}{~{\rm \ldots~per~}$b${\rm -tagged~Jet~Count}~} \\ \hline
$1$ & $20$ & $6$ & $15$ \\ \hline
$2$ & $12$ & $2$ & $6$ \\ \hline
$3$ & $3$ & $0$ & $1$ \\ \hline
$4$ & $1$ & $0$ & $0$ \\ \hline
\hline \multicolumn{4}{|c|}{~{\rm \ldots~per~Net~Isolated~Lepton~Count}~} \\ \hline
$1$ & $20$ & $6$ & $15$ \\ \hline
$2$ & $6$ & $1$ & $3$ \\ \hline
$3$ & $1$ & $0$ & $1$ \\ \hline
\hline \multicolumn{4}{|c|}{~{\rm \ldots~per~Isolated}~$\tau${\rm ~Count}~{\it with}~$\ge~2~b${\rm -tagged~Jets}~} \\ \hline
$1$ & $6$ & $0$ & $3$ \\ \hline
$2$ & $2$ & $0$ & $0$ \\ \hline
		\end{tabular}
		\label{tab:rawcms}
\end{table}

\begin{table}[htbp]
  \small
	\centering
	\caption{Distributed integral event counts for $1~{\rm fb}^{-1}$ and $\sqrt{s}=7$~TeV for the baseline ultra-high jet cut criteria.}
		\begin{tabular}{|c||c|c|c|} \hline
\multicolumn{4}{|c|}{\rm Raw Ultra-High Jet Event Distribution} \\ \hline
~~Multiplicity~~ & ~~${\cal F}$-$SU(5)$~~ & ~~SPS~SP3~~ & ~~$t\overline{t}+{\rm jets}$~~ \\ \hline
\hline \multicolumn{4}{|c|}{~{\rm Surviving~Events~per~Jet~Count}~} \\ \hline
$9$ & $38$ & $2$ & $8$ \\ \hline
$10$ & $26$ & $1$ & $1$ \\ \hline
$11$ & $16$ & $0$ & $1$ \\ \hline
$12$ & $8$ & $0$ & $0$ \\ \hline
$13$ & $4$ & $0$ & $0$ \\ \hline
$14$ & $1$ & $0$ & $0$ \\ \hline
$~{\rm Net}~$ & $93$ & $3$ & $10$ \\ \hline
\hline \multicolumn{4}{|c|}{~{\rm \ldots~per~}$b${\rm -tagged~Jet~Count}~} \\ \hline
$1$ & $23$ & $1$ & $5$ \\ \hline
$2$ & $27$ & $1$ & $4$ \\ \hline
$3$ & $18$ & $0$ & $0$ \\ \hline
$4$ & $9$ & $0$ & $0$ \\ \hline
$5$ & $3$ & $0$ & $0$ \\ \hline
$6$ & $1$ & $0$ & $0$ \\ \hline
\hline \multicolumn{4}{|c|}{~{\rm \ldots~per~Net~Isolated~Lepton~Count}~} \\ \hline
$1$ & $26$ & $1$ & $3$ \\ \hline
$2$ & $5$ & $0$ & $1$ \\ \hline
$3$ & $1$ & $0$ & $0$ \\ \hline
\hline \multicolumn{4}{|c|}{~{\rm \ldots~per~Isolated}~$\tau${\rm ~Count}~{\it with}~$\ge~2~b${\rm -tagged~Jets}~} \\ \hline
$1$ & $15$ & $0$ & $1$ \\ \hline
$2$ & $3$ & $0$ & $0$ \\ \hline
		\end{tabular}
		\label{tab:rawultrahigh}
\end{table}

We conclude our analysis with a look at the applicability to ultra-high jet events of two leading indicators of false
missing energy signatures, namely $\alpha_{\rm T}$ and $\Delta \phi^*$, as introduced in Section~\ref{sct:cuts}.
Since we have set the hard and soft jet thresholds identically in the ultra-high jet selection criteria, the similarly purposed
ratio $R(H^{\rm miss}_{\rm T})$ will be identically equal to one;  We therefore forgo any further discussion of this statistic.

\begin{figure}[htbp]
        \centering
        \includegraphics[width=0.45\textwidth]{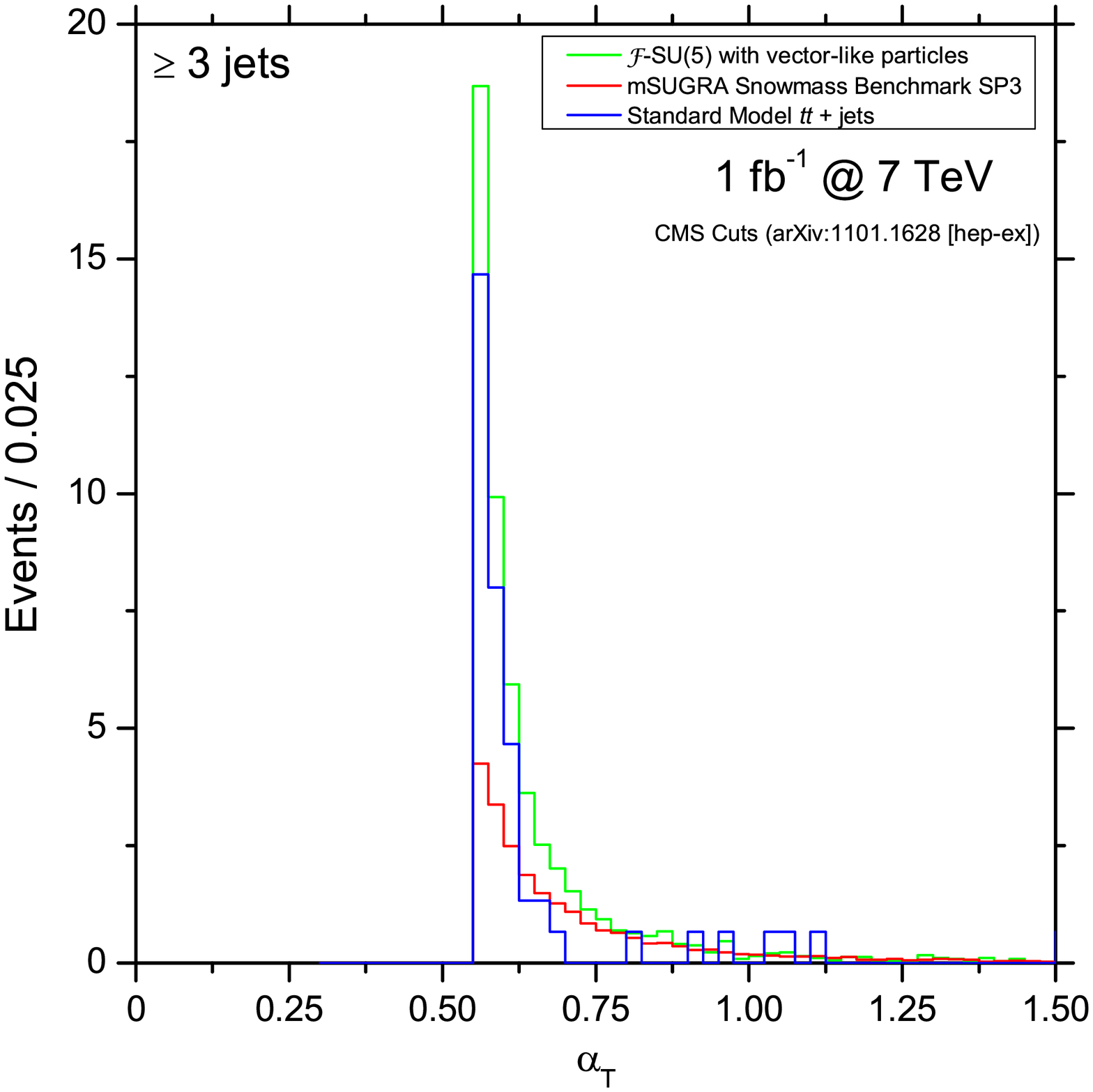}
        \caption{$\alpha_{\rm T}$ for events with $\ge$ 3 jets for 1 ${\rm fb}^{-1}$ and $\sqrt{s}$ = 7~TeV. Our emulation of the CMS style cuts is employed.}
\label{fig:gte3jets_alphaT}
\end{figure}

\begin{figure}[htbp]
        \centering
        \includegraphics[width=0.45\textwidth]{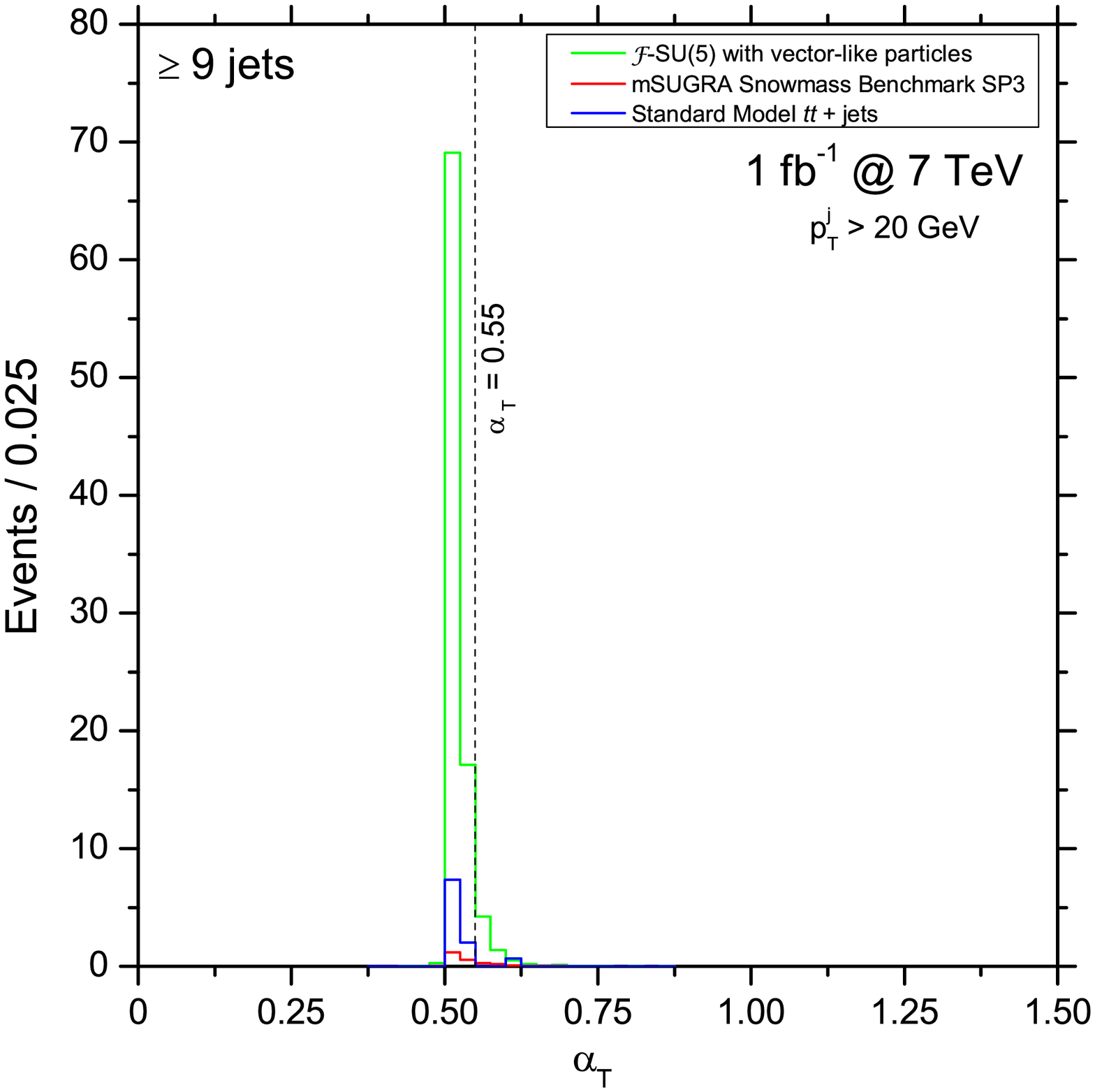}
        \caption{$\alpha_{\rm T}$ for events with $\ge$ 9 jets for 1 ${\rm fb}^{-1}$ and $\sqrt{s}$ = 7~TeV. Minimum $p_{\rm T}$ for a single jet is 20 GeV.}
\label{fig:gte9jets_alphaT}
\end{figure}

\begin{figure}[htbp]
        \centering
        \includegraphics[width=0.45\textwidth]{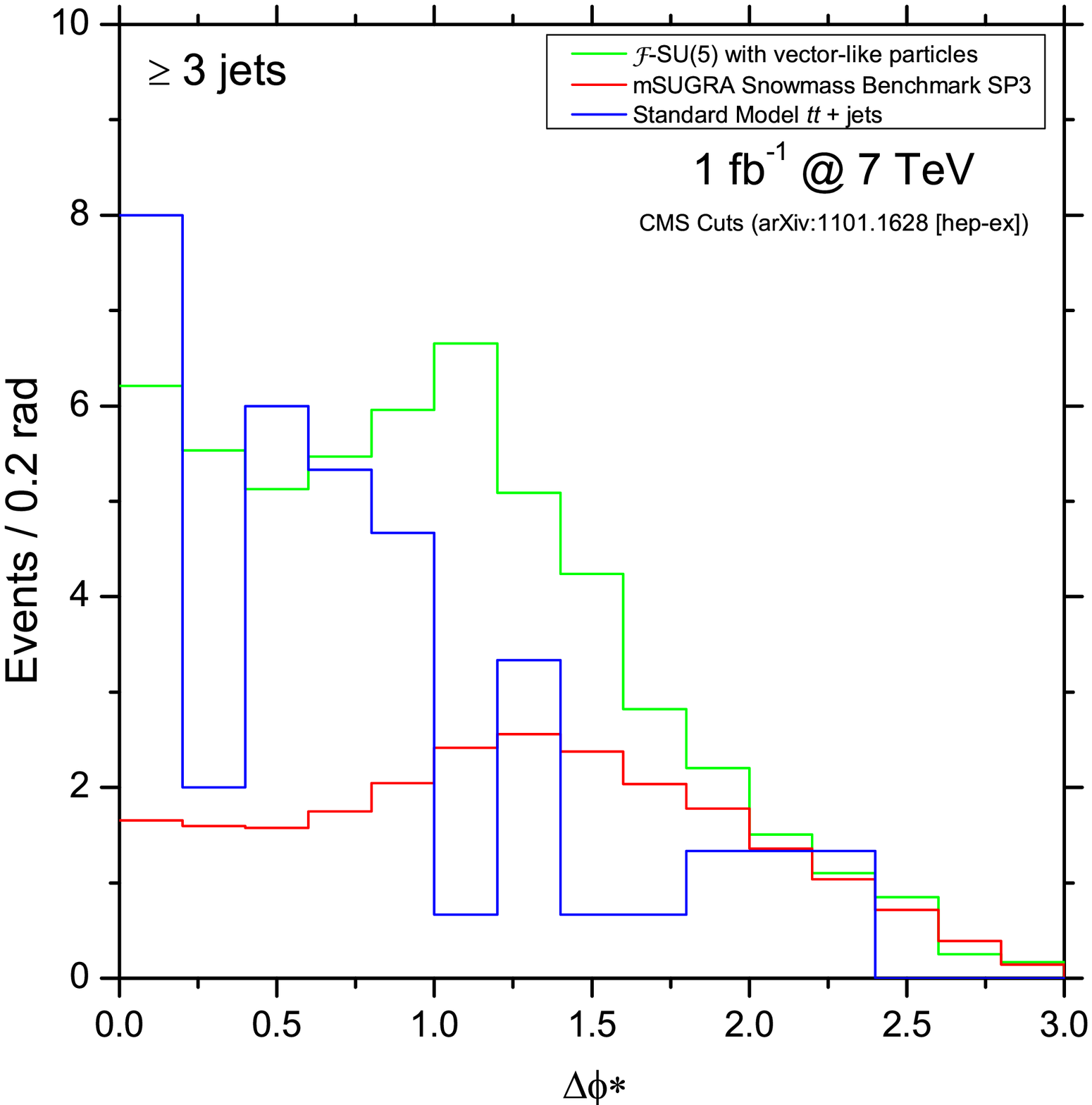}
        \caption{$\Delta \phi^*$ for events with $\ge$ 3 jets for 1 ${\rm fb}^{-1}$ and $\sqrt{s}$ = 7~TeV.  Our emulation of the CMS style cuts is employed.}
\label{fig:gte3jets_deltaphi}
\end{figure}

\begin{figure}[htbp]
        \centering
        \includegraphics[width=0.45\textwidth]{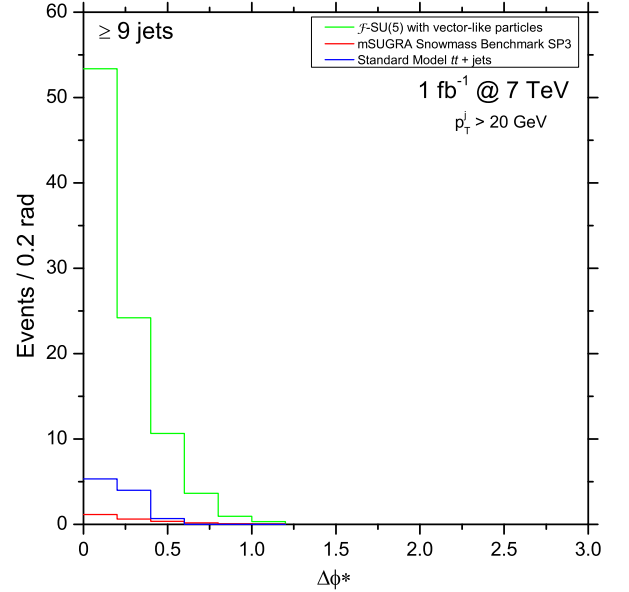}
        \caption{$\Delta \phi^*$ for events with $\ge$ 9 jets for 1 ${\rm fb}^{-1}$ and $\sqrt{s}$ = 7~TeV. Minimum $p_{\rm T}$ for a single jet is 20 GeV.}
\label{fig:gte9jets_deltaphi}
\end{figure}

Figs.~(\ref{fig:gte3jets_alphaT} - \ref{fig:gte9jets_deltaphi}) depict histograms of event counts dimensionlessly binned at
intervals of $0.025$ in $\alpha_{\rm T}$, and $0.2$ radians in $\Delta \phi^*$, for each of the CMS style and baseline
($p_{\rm T} > 20$, ${\rm jets} \ge 9$) ultra-high jet cuts.
Of course, it should be remarked again that for the CMS styled $\ge 3$ jet selection criteria, we are
not justified in reducing the SM background to only the $t \overline{t} + {\rm jets}$ constituent
processes.  Compare against the backgrounds borrowed directly from the CMS collaboration in
Fig.~(\ref{fig:FSU5_SP3_CMS}) for a visual estimate of the enhanced participation of the
QCD multi-jet and $W,Z + {\rm jets}$ processes in this context.

It is worth recalling here that the $\alpha_{\rm T}$ statistic was originally devised for di-jet processes,
and later adapted to multi-jet events by the artful assemblage of two optimized pseudo-jets from the
full set of tracks.  Its intrinsic relevance for the scaling up to ultra-high jet processes may then be held in some doubt.
Indeed, Fig.~(\ref{fig:gte3jets_alphaT}), demonstrates a healthy tail of events which survive the hard CMS $\alpha_{\rm T} \ge 0.55$
cut, while the surviving ultra-high jet events of Fig.~(\ref{fig:gte9jets_alphaT}), for which no $\alpha_{\rm T}$ cut is imposed,
cluster very tightly about the value of $1/2$, and would fail, almost {\it en masse}, a restriction to $\alpha_{\rm T} \ge 0.55$.
Apparently, the large multiplicity of available, relatively soft, jets makes it quite likely that a reasonably well balanced
pair of pseudo-jets may be constructed.

The biased $\Delta \phi^*$ statistic appears bound to face a similar outcome.  Fig.~(\ref{fig:gte3jets_deltaphi}), depicting the CMS style cuts,
reveals a reasonably well balanced angular distribution, with the greatest bias toward zero occurring for the SM contributions, which are
indeed expected to be primary culprits in the counterfeit of missing energy.  Although $\Delta \phi^*$ is not expressly activated in
either of our Table~\ref{tab:cuts} criteria, it remains a statistic of common use and relevance for intermediate jet multiplicity applications.
By contrast, Fig.~(\ref{fig:gte9jets_deltaphi}) shows all surviving ultra-high jet events to cluster very closely to the $\Delta \phi^* = 0$ home base.
The reason, again, seems to be that with so many constituent jets available for analysis, it becomes quite likely that the angular orientation of at least one jet
might be sufficiently well azimuthally aligned with the true missing energy track that its rescaling could rebalance the event. 

It seems that these two common cuts, so beneficial for reduction of the SM background against intermediate jet multiplicity events, will not only
fail to efficiently differentiate ultra-high jet multiplicities, but will moreover preferentially indicate against an ultra-high jet signal.
We take this as further justification for the exclusion of both $\alpha_{\rm T}$ and $\Delta \phi^*$ from our optimized ultra-high jet multiplicity
search strategy, and emphasize again that the ultra-high jet blockade itself forms a sufficiently strong discriminant against both the SM and
typical mSUGRA attempts at a post-SM solutions.  The $\cal{F}$-$SU(5)$ signature represents therefore a clearly defined case study of a realistic
SUSY signal which, although readily discoverable in principle, would be severely attenuated, and potentially concealed,
by the data selection cuts standard to the most prominent CMSSM studies.

%%%%%%%%%%%%%%%%%%%%%%%%%%%%%%%%%%%%%%%%%%%%%%%%%%%%%%%%%%%%%%%%%%%%%%%%%%%%

\section{Conclusions}

The LHC era has been long anticipated, with expectations for the revelation of physics beyond the Standard Model
mounting ever higher as the first forays into this new high energy frontier begin finally to return preliminary
experimental results.  With the prospects for hard evidentiary insight into the structure of the underlying next-generation
theory enticingly close at hand, the field of prospective SUSY models and their respective LHC fingerprints has grown
substantially. Nevertheless, our exploration of recently published signatures for supersymmetry discovery reveals a
common focus toward low-multiplicity jet events or lepton rich events, owing much to the parameter space of the CMSSM.

We have showed here that an analysis of LHC data which is skewed toward these low-multiplicity jet events could
mask an authentic and potentially well resolved SUSY signal which bears a sufficiently distinct signature.
By no means pretending a special knowledge of the true theory, we are nonetheless convinced that the efficiency and
manifold phenomenological merits of the model named No-Scale ${\cal F}$-$SU(5)$ justify a comprehensive attempt
at falsification of its characteristic collider level predictions. The achievements of this model are noteworthy
indeed: the experimentally viable parameter space is condensed to a single point for fixed vector-like mass, likewise
constrained to a string of points for all vector-like mass, and non-trivially consistent with the dynamic theoretical
determination of $\tan \beta$ and the single modulus parameter $M_{1/2}$. 

We now append to these attainments a clear and convincing ultra-high jet multiplicity signal for events
with at least nine jets, unmistakable for the SM background or the CMSSM.  The optimized post-processing selection
cuts which have been outlined in this work are essential for the discovery of supersymmetry if No-Scale ${\cal F}$-$SU(5)$ is
indeed highly proximal to the physical model.  We have released our statistics processing and event cutting program
{\tt CutLHCO} into the public domain along with this publication.  Our suggestions for an alternate cutting analysis do not
constitute severe deviations from the spirit of existing cut methodologies, with the two chief adjustments being reduction of
the minimum transverse momentum $p_{\rm T}$ per jet, and an escalation of the minimum number of jets in a subscribed event.
However elementary these modifications may seem, the consequences could be considerable.

We have presented a detailed Monte Carlo simulation analysis of the early LHC run at an energy of ${\sqrt s}=7$ TeV and an integrated
luminosity of $1~{\rm fb}^{-1}$, for the leading SM background, one mSUGRA scenario, and a fully representative updated benchmark
of the No-Scale $\cal{F}$-$SU(5)$ model, cutting at $p_{\rm T} > 10$~GeV and $p_{\rm T} > 20$~GeV in turn, for clusters of
$\ge 9$ or $\ge 10$, and $\ge 11$ or $\ge 12$ jets.  We showed that the No-Scale $\cal{F}$-$SU(5)$ scenario can be clearly
distinguished from the SM background and the mSUGRA scenario, and can be tested at the early LHC run by the end of 2011.
Moreover, we pointed out that an essential uniformity, modulo an overall rescaling, of the viable parameter space suggests that
the entirety of No-Scale ${\cal F}$-$SU(5)$ may be testable by the end of 2012. 

Detection of such a signal of stringy origin by the LHC could reveal not just the flipped nature of the
high-energy theory, but also shed light on the geometry of the hidden compactified six-dimensional manifold.
Thus, the stakes could not be higher for potential identification of the ultra-high jet events or the revelations more profound.

%%%%%%%%%%%%%%%%%%%%%%%%%%%%%%%%%%%%%%%%%%%%%%%%%%%%%%%%%%%%%%%%%%%%%%%%%%%%

\begin{acknowledgments}
This research was supported in part 
by  the DOE grant DE-FG03-95-Er-40917 (TL and DVN),
by the Natural Science Foundation of China 
under grant numbers 10821504 and 11075194 (TL),
and by the Mitchell-Heep Chair in High Energy Physics (JM).
We thank Alexei N. Safonov for helpful discussions.
We thank Sam Houston State University for providing high performance computing resources.
\end{acknowledgments}

%%%%%%%%%%%%%%%%%%%%%%%%%%%%%%%%%%%%%%%%%%%%%%%%%%%%%%%%%%%%%%%%%%%%%%%%%%%%

\appendix*

\section{No-Scale $\cal{F}$-$SU(5)$\label{app:fsu5}}

\subsection{Phenomenological Overview}

We have recently demonstrated~\cite{Li:2010ws,Li:2010mi} the unique phenomenological
consistency and profound predictive capacity of a model dubbed
No-Scale ${\cal F}$-$SU(5)$, resting essentially and in equal measure on the tripodal
foundations of the ${\cal F}$-lipped $SU(5)$ Grand Unified Theory
(GUT)~\cite{Barr:1981qv,Derendinger:1983aj,Antoniadis:1987dx}, two pairs of
hypothetical TeV scale vector-like supersymmetric multiplets with origins in
${\cal F}$-theory model building~\cite{Jiang:2006hf,Jiang:2009zza,Jiang:2009za,Li:2010dp,Li:2010rz},
and the dynamically established boundary conditions of No-Scale Supergravity
(SUGRA)~\cite{Cremmer:1983bf,Ellis:1983sf, Ellis:1983ei, Ellis:1984bm, Lahanas:1986uc}.
It appears that the No-Scale scenario, and most stringently the vanishing of the Higgs
bilinear soft term $B_\mu$, comes into its own only when applied at an elevated
scale, approaching the Planck mass.  $M_{\cal F}$, the point of the ultimate second stage
$SU(5)\times U(1)_{\rm X}$ unification, emerges in turn as a suitable candidate scale
only when substantially decoupled from the penultimate GUT scale unification
of $SU(3)_C\times SU(2)_L$ at $M_{32} \simeq 10^{16}$~GeV via the modification to
the renormalization group equations (RGEs) from the extra vector-like multiplets.

We have systematically established the hyper-surface within
the $\tan \beta$, top quark mass $m_{t}$, gaugino mass
$M_{1/2}$, and vector-like particle mass $M_{V}$ parameter
volume which is compatible with the application of the simplest
No-Scale SUGRA boundary conditions~\cite{Cremmer:1983bf,Ellis:1983sf, Ellis:1983ei, Ellis:1984bm, Lahanas:1986uc}.
We have demonstrated that simultaneous adherence to all current experimental
constraints, most importantly contributions to the muon anomalous
magnetic moment $(g-2)_\mu$~\cite{Bennett:2004pv}, the branching ratio limit on
$(b \rightarrow s\gamma)$~\cite{Barberio:2007cr, Misiak:2006zs},
and the 7-year WMAP relic density measurement~\cite{Komatsu:2010fb},
dramatically reduces the allowed solutions to a highly non-trivial
``golden strip'', tightly confining $\tan \beta$, $m_{t}$, $M_{1/2}$, and $M_{V}$, effectively
eliminating all extraneously tunable model parameters, where the consonance of the
theoretically viable $m_{t}$ range with the experimentally established value~\cite{:2009ec}
may be interpreted an independently correlated ``postdiction''.  Finally, taking a fixed $Z$-boson mass,
we have dynamically determined the universal gaugino mass $M_{1/2}$ and fixed $\tan \beta$ via the ``Super No-Scale''
mechanism~\cite{Li:2010uu}, that being the secondary minimization, at a local {\it minimum minimorum},
of the minimum $V_{\rm min}$ of the Higgs potential for the electroweak symmetry breaking (EWSB) vacuum.

This model is moreover quite
interesting from a phenomenological point of view~\cite{Jiang:2009zza,Jiang:2009za}. The predicted
vector-like particles can be observed at the Large Hadron Collider (LHC), though possibly
not during the initial run.  The partial lifetime for proton decay
in the leading ${(e|\mu)}^{+} \pi^0 $ channels falls around
$5 \times 10^{34}$ years~\cite{Li:2010dp,Li:2010rz}, testable at the future
Hyper-Kamiokande~\cite{Nakamura:2003hk} and
Deep Underground Science and Engineering Laboratory (DUSEL)~\cite{Raby:2008pd}
experiments~\cite{Li:2009fq, Li:2010dp, Li:2010rz}.
The lightest CP-even Higgs boson mass can be increased~\cite{HLNT},
hybrid inflation can be naturally realized, and the
correct cosmic primordial density fluctuations can be
generated~\cite{Kyae:2005nv}.

\subsection{The $\cal{F}$-lipped SU(5) GUT}

Gauge coupling unification strongly suggests the existence of a GUT.
In minimal supersymmetric $SU(5)$ models
there are problems with doublet-triplet splitting and dimension
five proton decay by colored Higgsino exchange~\cite{Antoniadis:1987dx}. These difficulties
can be elegantly overcome in Flipped $SU(5)$ GUT
models~\cite{Barr:1981qv, Derendinger:1983aj, Antoniadis:1987dx}
via the missing partner mechanism~\cite{Antoniadis:1987dx}.

Written in full, the gauge group of Flipped $SU(5)$ is
$SU(5)\times U(1)_{X}$, which can be embedded into $SO(10)$.
The generator $U(1)_{Y'}$ is defined for fundamental five-plets as
$-1/3$ for the triplet members, and $+1/2$ for the doublet.
The hypercharge is given by $Q_{Y}=( Q_{X}-Q_{Y'})/5$.
There are three families of Standard Model (SM) fermions,
whose quantum numbers under the $SU(5)\times U(1)_{X}$ gauge group are
\begin{eqnarray}
F_i={\mathbf{(10, 1)}},\quad {\bar f}_i={\mathbf{(\bar 5, -3)}},\quad
{\bar l}_i={\mathbf{(1, 5)}},
\label{smfermions}
\end{eqnarray}
where $i=1, 2, 3$. 
To break the GUT and electroweak gauge symmetries, we 
introduce two pairs of Higgs fields:
a pair of ten-plet Higgs for breaking the GUT symmetry, and a pair
of five-plet Higgs for electroweak symmetry breaking. 
\begin{eqnarray}
& H={\mathbf{(10, 1)}}\quad;\quad~{\overline{H}}={\mathbf{({\overline{10}}, -1)}} &\\
& h={\mathbf{(5, -2)}}\quad;\quad~{\overline h}={\mathbf{({\bar {5}}, 2)}} &
\label{Higgse1}
\end{eqnarray}

A most notable intrinsic feature of the Flipped $SU(5)$ GUT is the presence of dual unification scales, with the
ultimate merger of $SU(5) \times U(1)_{\rm X}$ at the scale $M_{\cal F}$ occurring subsequent in energy to the penultimate
$SU(3)_{\rm C}$ and $SU(2)_{\rm L}$ mixing at $M_{32}$.  In the more traditional Flipped $SU(5)$ formulations, $M_{\cal F}$ has been only
slightly elevated from $M_{32}$, larger by a factor of perhaps only two or three~\cite{Ellis:2002vk}.  Our interest
however, is in scenarios where the ratio $M_{\cal F}/M_{32}$ is considerably larger, on the order of $10$ to $100$.

Key motivations for this picture include the desire to address the monopole problem via hybrid inflation,
and the opportunity for realizing true string scale gauge coupling unification in
the free fermionic model building context~\cite{Jiang:2006hf, Lopez:1992kg},
or the decoupling scenario in F-theory models~\cite{Jiang:2009zza,Jiang:2009za}.
We have previously also considered the favorable effect of such considerations on the decay rate of the proton~\cite{Li:2010dp,Li:2010rz}.
Our greatest present interest however, is the effortless manner in which the elevation of the $SU(5) \times U(1)_{\rm X}$
scale salvages the dynamically established boundary conditions of No-Scale Supergravity.  Being highly predictive,
these conditions are thus also intrinsically highly constrained, and notoriously difficult to realize generically.

\subsection{$\cal{F}$-theory Vector-Like Multiplets}

We have introduced additional vector-like particle multiplets derived
within the $\cal{F}$-theory~\cite{Jiang:2006hf} model building context
to address the ``little hierarchy'' problem, altering the $\beta$-coefficients
of the renormalization group to dynamically elevate the secondary 
$SU(5)\times U(1)_{\rm X}$ unification at $M_{\cal F}$ to near the Planck
scale, while leaving the $SU(3)_C\times SU(2)_L$ unification at $M_{32}$
close to the traditional GUT scale.  In other words,
one obtains true string-scale gauge coupling unification in 
free fermionic string models~\cite{Jiang:2006hf,Lopez:1992kg} or
the decoupling scenario in F-theory models~\cite{Jiang:2009zza,Jiang:2009za}.
To avoid a Landau pole for the strong
coupling constant, we are restricted around the TeV scale
to one of the following two multiplet sets~\cite{Jiang:2006hf}.
\begin{eqnarray}
\hspace{-.3in}
& Z1:~ \left( {XF}_{\mathbf{(10,1)}} \equiv (XQ,XD^c,XN^c),~{\overline{XF}}_{\mathbf{({\overline{10}},-1)}} \right)& \nonumber \\
&  Z2:~\left( {XF}, ~{\overline{XF}},
{Xl}_{\mathbf{(1, -5)}},~{\overline{Xl}}_{\mathbf{(1, 5)}}\equiv XE^c \right) &\label{z1z2}
\end{eqnarray}
In the prior, $XQ$, $XD^c$, $XE^c$, $XN^c$ have the same quantum numbers as the
quark doublet, the right-handed down-type quark, charged lepton, and
neutrino, respectively.  We have argued~\cite{Li:2010mi} that the
feasibly near-term detectability of these hypothetical fields in collider experiments,
coupled with the distinctive flipped charge assignments within the multiplet structure,
represents a smoking gun signature for Flipped $SU(5)$, and have thus coined the term
{\it flippons} to collectively describe them.
In this paper, we consider only the $Z2$ set, although discussion for the $Z1$ set,
if supplemented by heavy threshold corrections, can be similar.

We emphasize that the specific representations of vector-like fields which we currently employ have been explicitly
constructed within the local F-theory model building context~\cite{Jiang:2009zza, Jiang:2009za}.  However, the mass of these fields,
and even the fact of their existence, is not mandated by the F-theory, wherein it is also possible to
realize models with only the traditional Flipped (or Standard) $SU(5)$ field content.  We claim only an inherent consistency of their conceptual
origin out of the F-theoretic construction, and take the manifest phenomenological benefits which accompany the elevation of
$M_{\cal F}$ as justification for the greater esteem which we hold for this particular model above other alternatives.

\subsection{No-Scale Supergravity}

The Higgs boson, being a Lorentz scalar,
is not stable in the SM against quadratic quantum mass corrections
which drive it toward the dominant Planck scale, some
seventeen orders of magnitude above the value required for consistent 
EWSB.  Supersymmetry naturally solves this fine tuning problem
by pairing the Higgs with a chiral spin-$1/2$ ``Higgsino'' partner field, and
following suit with a corresponding bosonic (fermionic) superpartner for all
fermionic (bosonic) SM fields, introducing the full set of quantum counter terms.
Localizing the supersymmetry (SUSY) algebra, which includes the generator of
spacetime translations (the momentum operator), induces general coordinate invariance,
producing the supergravity (SUGRA) theories.

Since we do not observe mass degenerate superpartners for the known SM fields,
SUSY must itself be broken around the TeV scale.
In the traditional framework, supersymmetry is broken in the hidden sector, and the effect is 
mediated to the observable sector via gravity or gauge interactions.
In GUTs with minimal gravity mediated supersymmetry breaking, called mSUGRA,
one can fully characterize the supersymmetry breaking
soft terms by four universal parameters, the
gaugino mass $M_{1/2}$, scalar mass $M_0$, trilinear coupling $A$, and
the low energy ratio $\tan \beta$ of up- to down-type Higgs VEVs,
plus the sign of the Higgs bilinear mass term $\mu$.
The $\mu$ term and its bilinear
soft term $B_{\mu}$ are determined
by the $Z$-boson mass $M_Z$ and $\tan \beta$ after
the electroweak (EW) symmetry breaking.

No-Scale Supergravity was proposed~\cite{Cremmer:1983bf,Ellis:1983sf, Ellis:1983ei, Ellis:1984bm, Lahanas:1986uc}
to address the cosmological flatness problem,
and defined as the subset of supergravity models
which satisfy the following three constraints~\cite{Cremmer:1983bf}:
(i) The vacuum energy vanishes automatically due to the suitable
K\"ahler potential; (ii) At the minimum of the scalar
potential, there are flat directions which leave the
gravitino mass $M_{3/2}$ undetermined; (iii) The super-trace
quantity ${\rm Str} {\cal M}^2$ is zero at the minimum. Without this,
the large one-loop corrections would force $M_{3/2}$ to be either
zero or of Planck scale.  The defining K\"ahler potential~\cite{Ellis:1984bm}
\begin{eqnarray}
K &=& -3 {\rm ln}( T+\overline{T}-\sum_i \overline{\Phi}_i
\Phi_i)
\label{NS-Kahler}
\end{eqnarray}
automatically satisfies the first two conditions, while
the third is model dependent and can always be satisfied in
principle~\cite{Ferrara:1994kg}.

In Eq.~(\ref{NS-Kahler}), $T$ is a modulus field, while the
$\Phi_i$ are $N_C$ scalar matter fields which parameterize the
coset space $SU(N_C+1, 1)/(SU(N_C+1)\times U(1))$.
The scalar potential is automatically positive semi-definite,
and has a flat direction along the $T$ field.
The non-compact structure of the symmetry implies that the classical vacuum
is not only constant but actually identical to zero.
Moreover, the simplest No-Scale boundary
conditions $M_0=A=B_{\mu}=0$ are dynamically established,
while $M_{1/2}>0$ is allowed, and indeed required for SUSY breaking.
The CP violation problem and the flavor changing neutral current
problems are automatically solved in turn.
All low energy scales are dynamically generated by quantum corrections,
{\it i.e.}~running under the RGEs, to the classically flat potential.

\subsection{The Stringy Super No-Scale Mechanism}

The fiercely reductionist No-Scale picture inherits
an associative weight of motivation from its robustly generic and natural appearance, for example,
in the compactification of the weakly coupled heterotic string theory~\cite{Witten:1985xb}, compactification of
M-theory on $S^1/Z_2$ at the leading order~\cite{Li:1997sk},
and potentially also directly in F-theory models~\cite{Beasley:2008dc,Beasley:2008kw, Donagi:2008ca, Donagi:2008kj}.

In the simplest stringy No-Scale SUGRA, the K\"ahler
modulus $T$, a characteristic of the Calabi-Yau manifold,
is the single relevant modulus field, the dilaton coupling being irrelevant.
The F-term of $T$ generates the gravitino mass $M_{3/2}$, 
which is proportionally equivalent to $M_{1/2}$.
Exploiting the simplest No-Scale boundary condition at $M_{\cal F}$ and 
running from high energy to low energy under the RGEs,
there can be a secondary minimization, or {\it minimum minimorum}, of the minimum of the
Higgs potential $V_{\rm min}$ for the EWSB vacuum.
Since $V_{\rm min}$ depends on $M_{1/2}$, the universal gaugino mass $M_{1/2}$ is consequently 
dynamically determined by the equation $dV_{\rm min}/dM_{1/2}=0$,
aptly referred to as the ``Super No-Scale'' mechanism;
We have argued by the combined action of this mechanism,
the transmutative role of the RGEs, and the stabilizing counter-balance of
supersymmetry, that No-Scale $\cal{F}$-$SU(5)$ addresses the various aspects of the
gauge hierarchy problem~\cite{Li:2010uu}.

The three parameters $M_0,A,B_{\mu}$ are once again identically zero at the
boundary because of the defining K\"ahler potential, and are thus known at all other scales as well by the RGEs.  The
minimization of the Higgs scalar potential with respect to the neutral elements of both SUSY Higgs doublets gives two
conditions, the first of which fixes the magnitude of $\mu$.  The second condition, which would traditionally be used
to fix $B_{\mu}$, instead here enforces a consistency relationship on the remaining parameters, being that
$B_{\mu}$ is already constrained.

In general, the $B_{\mu} = 0$ condition gives a hypersurface of solutions cut out from a very large parameter space.
If we lock all but one parameter, it will give the final value.  If we take a slice of two dimensional space, as has been 
described, it will give a relation between two parameters for all others fixed.
In a three-dimensional view with $B_{\mu}$ on the vertical axis, this
curve is the ``flat direction'' line along the bottom of the trench of $B_{\mu}=0$ solutions.  In general, we
must vary at least two parameters rather than just one in isolation, in order that their mutual compensation may transport
the solution along this curve.

It must be emphasized that the $B_{\mu}=0$ No-Scale
boundary condition is the central agent affording this determination, as it is the extraction of the parameterized
parabolic curve of solutions in the two compensating variables which allows for a localized, bound nadir point to be
isolated by the Super No-Scale condition, dynamically determining {\it both} parameters.  The background surface of
$V_{\rm min}$ for the full parameter space outside the viable $B_{\mu}=0$ subset is, in contrast, a steadily inclined
and uninteresting function.  We have demonstrated that the local {\it minimum minimorum} of $V_{\rm min}$ for
selected inputs of $M_{V}$ and $m_{t}$ may be taken to dynamically establish the values of the pair of prominent unknown
inputs $M_{1/2}$ and $\tan \beta$~\cite{Li:2010uu}.
Although $M_{1/2}$ and $\tan \beta$ have no {\it directly} established experimental values, they are severely indirectly constrained by
phenomenology in the context of this model~\cite{Li:2010ws,Li:2010mi}.  It is highly non-trivial that there should be
a strong accord between the top-down and bottom-up perspectives, but this is indeed precisely what has been observed~\cite{Li:2010uu}. 

%%%%%%%%%%%%%%%%%%%%%%%%%%%%%%%%%%%%%%%%%%%%%%%%%%%%%%%%%%%%%%%%%%%%%%%%%%%%

\bibliography{bibliography}

\end{document}